\documentclass[a4paper,11pt]{article}
\pdfoutput=1 % if your are submitting a pdflatex (i.e. if you have
\usepackage{jinstpub}
\usepackage{lineno}
\usepackage{adjustbox}
\usepackage{multirow}
\title{DANTE Digital Pulse Processor for XRF and XAS experiments}

\author[a,1]{F.J.~Iguaz,\note{Corresponding author.}}
\author[b]{L.~Bombelli,}
\author[b]{S.~Meo,}
\author[a]{F.~Orsini,}
\author[a]{S.~Sch\"oeder,}
\author[b]{A.~Tocchio,}
\author[a]{N.~Trcera,}
\author[a]{D.~Vantelon}

\affiliation[a]{SOLEIL Synchrotron, L'Orme des Merisiers, D\'epartementale 128, 91190 Saint-Aubin, France}
\affiliation[b]{XGLAB s.r.l., via Conte Rosso 23, Milano, Italy}

\emailAdd{francisco-jose.iguaz-gutierrez@synchrotron-soleil.fr}

\abstract{DANTE is a new Digital Pulse Processor (DPP) developed for fluorescence detectors, like silicon drift detectors (SDDs) or High Purity Germanium detectors (HPGe), used in X-ray Fluorescence (XRF) and X-ray Absorption Spectroscopy (XAS) experiments at synchrotron facilities. Its main features are its optimal energy resolution and peak stability for detector count rate values up to 1-2~Mcps, and its enhanced rejection of pile-up events.
In this paper, we present the first complete evaluation of DANTE performance in SOLEIL synchrotron facility. DANTE has been tested in laboratory with an X-ray generator source and in different experiments at LUCIA and PUMA beamlines at SOLEIL.}


\arxivnumber{1234.56789} % only if you have one

\begin{document}
\maketitle
\flushbottom

%%%% Introduction
\section{Introduction}
\label{sec:intro}
\begin{figure}[htb!]
\centering
\includegraphics[width=140mm]{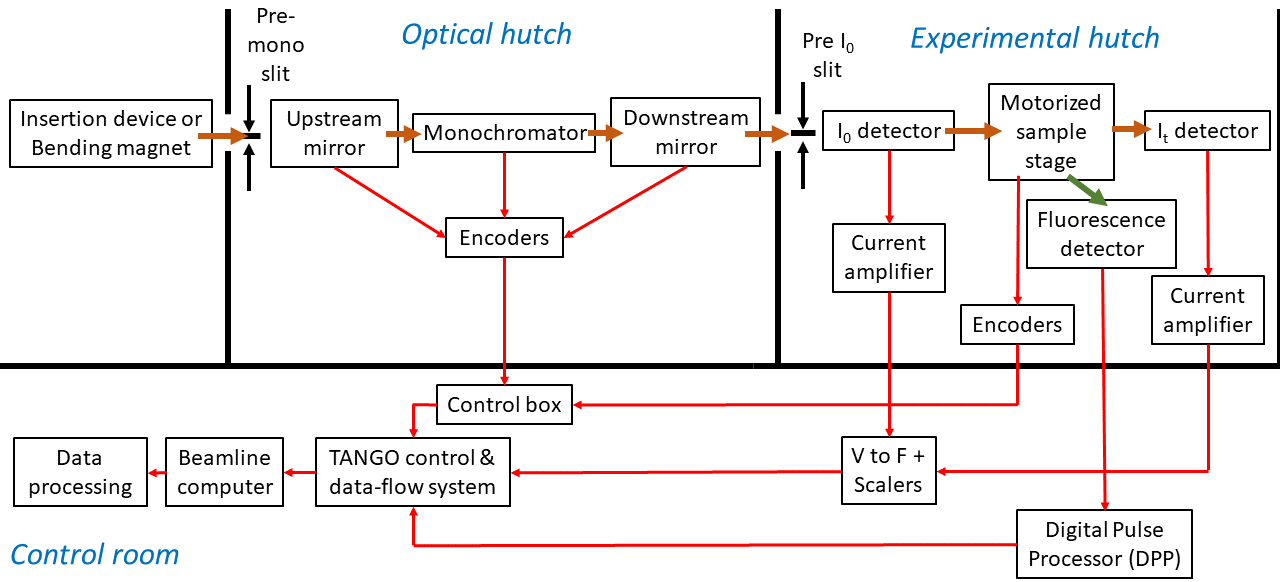}
\caption{General layout of a beamline using either an XRF or XAS technique at SOLEIL synchrotron, adapted from Fig.~5.2 of~\cite{calvin2013xafs}.}
\label{fig:BeamlineSchema}
\end{figure}

X-ray Fluorescence (XRF)~\cite{grieken2001xrf} and X-ray Absorption Spectrocopy (XAS)~\cite{calvin2013xafs} are two complementary techniques commonly used at synchrotron facilities. The first one allows to obtain the elemental composition of a heterogeneous sample, while the second one is a powerful tool to reveal the electronic and short range order structure of a given element in a heterogeneous matrix. Beamlines at SOLEIL synchrotron using any of these two techniques follow the scheme shown in Fig.~\ref{fig:BeamlineSchema}.
 
In an XRF experiment, the fluorescence signal of a sample is measured after an excitation by incident X-ray photons at a given energy. The atoms present in the sample are excited by the incident X-rays and, by a de-excitation process, each atom in the sample emits X-ray photons. In fluorescence mode, these photons are collected by an energy-resolved detector, either a Silicon Drift Detector (SDD)~\cite{SCHLOSSER2010270} or a High Purity Germanium detector (HPGe)~\cite{SANGSINGKEOW2003183}) equipped with a Digital Pulse Processor (DPP)~\cite{Bordessoule2019}. This signal is directly proportional to the element concentration in the sample and is calculated by the integral of each peak present in the fluorescence spectrum. Combined to a micro-sized X-ray beam, the $\mu$-XRF allows to collect elemental maps of a heterogeneous sample by scanning the sample position with fixed incident energy photons.

In an XAS experiment, the absorption coefficient of a chemical element is measured as a function of the scan in energy of the incident X-rays around the ionization threshold of the element. Using fluorescence mode for instance, the XAS spectra are obtained by the calculation of $I_f/I_0$ as a function of the incident energy. $I_f$ is the fluorescence signal emitted by the probed element and proportional to the absorption coefficient of the element, while $I_0$ is the intensity of the incident X-ray beam.

For the two techniques, the signal-to-noise (S/N) ratio of the experiment~\cite{Heald:rv5031} is determined by:
\begin{enumerate}
    \item The fluorescence intensity, which is limited by the maximum detector counting rate, and is of about $10^5$ to $10^6$~cps. This limit is defined either by the linear dependence range of Output Count Rate (OCR) with Input Count Rate (ICR), corresponding to a detector dead time of $\sim$10\%; or by the rate for which the detector stops being energy-resolved, at dead time values of $\sim$60\%.
    \item The background level, which is due to elastic or inelastic beam scattering, escape peaks from other elements in the sample or pile-up, i.e., two photons arriving very closely together in time and being counted as one. The background level also depends on the detector ICR.
\end{enumerate}

When either the dead time or the background level are too high, scientists are forced to restrict the detector solid angle to improve S/N ratio, either moving away the detector from the sample or using other techniques like collimators, low energy filters or soller slits.

A new generation of DPPs, like XIA-FalconX~\cite{XIAFalcon,Scoullar2011} or Xspress3~\cite{QuantumXspress3,FARROW1995567}, has provided better performance at high ICR by reducing the dead time, especially for Xspress3, and pile-up intensity, especially for XIA-FalconX~\cite{Bordessoule2019}. Another interesting DPP is now available commercially, DANTE, which features better performance than previous generation~\cite{XGLabDANTE}. This paper presents a first complete evaluation of DANTE performance in a synchrotron facility.
Other tests in synchrotron beamlines have been reported in~\cite{Bombelli2019TowardsOX,Pouyet2021ep}. This evaluation has been done in the laboratory with an X-ray generator source and in different XFS and XRF experiments at LUCIA and PUMA beamlines of SOLEIL synchrotron. Beamline tests were possible thanks to the integration of DANTE API library in SOLEIL TANGO control system~\cite{TANGO1999}.

In Sec.~\ref{sec:descDANTE}, a short technical description of DANTE DPP is made. In Sec.~\ref{sec:labmes}, we present the performance results of DANTE in laboratory with an X-ray generator source, including a comparison with those of recent DPPs. In Sec.~\ref{sec:LUCIAint}, we discuss the XRF and XAS experiments at LUCIA beamline, while Sec.~\ref{sec:PUMAint} is devoted to the XRF measurements at PUMA beamline. The conclusions (in Sec.~\ref{sec:conc}) complete this work.

\section{DANTE Digital Pulse Processor}
\label{sec:descDANTE}
\begin{figure}[htb!]
\centering
\includegraphics[width=50mm]{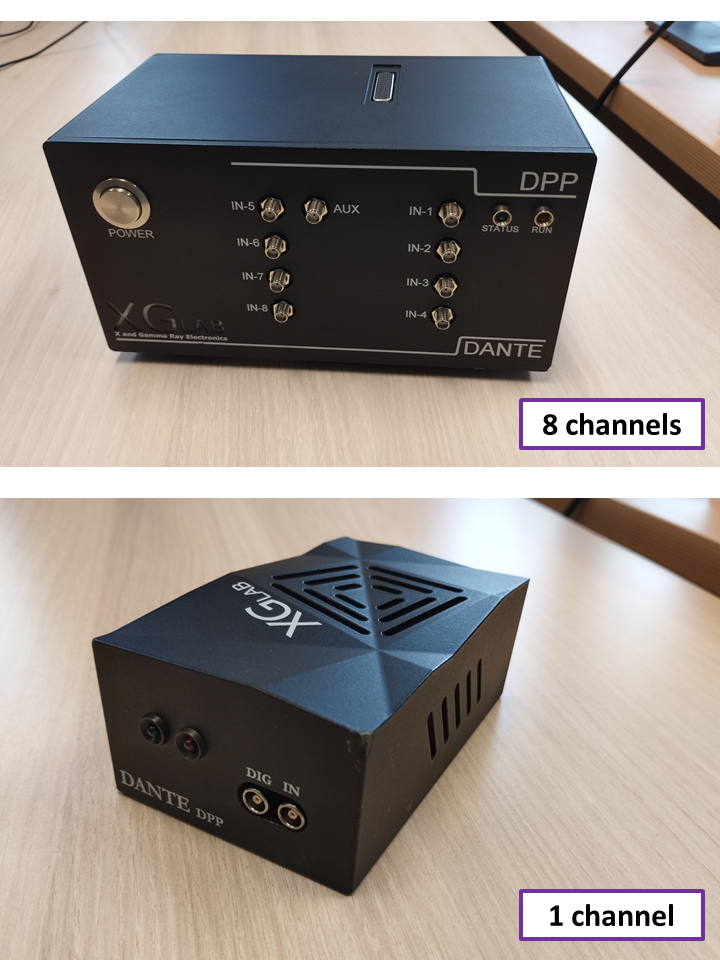}
\includegraphics[width=100mm]{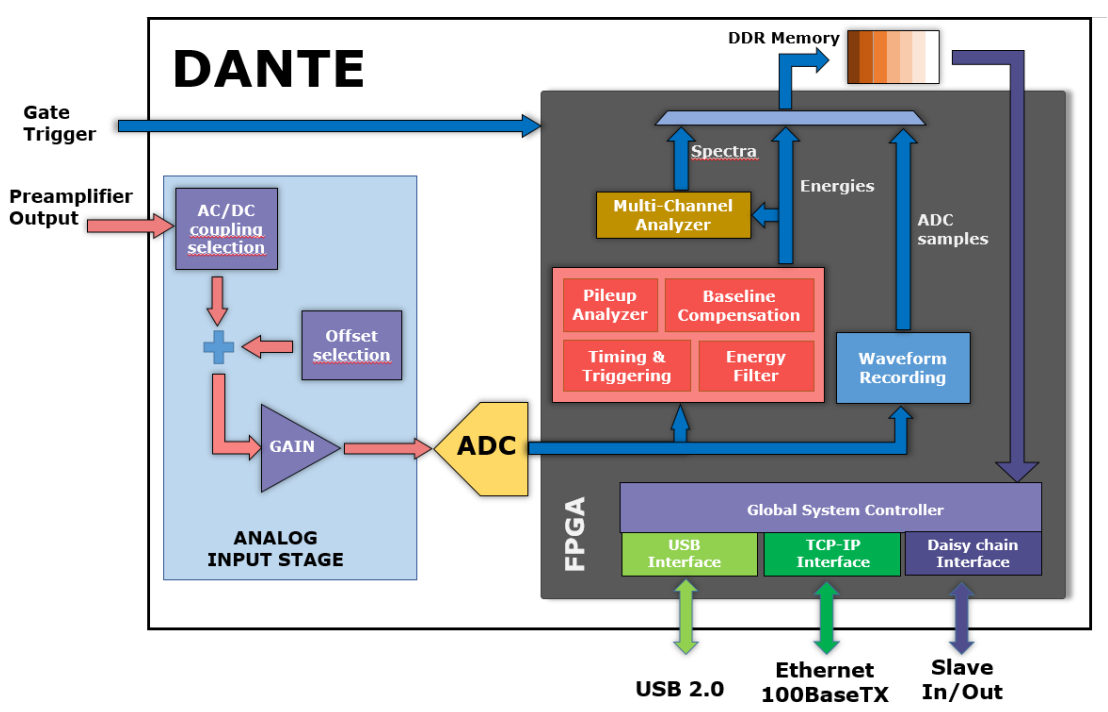}
\caption{Left: DANTE box in 1 and 8 channel version. Right: DANTE board scheme.}
\label{fig:DANTEScheme}
\end{figure}

DANTE DPP has been specifically designed for X-ray spectroscopy applications, like ultra-fast XRF elemental mapping, low-energy XRF and XAS. It has been designed to provide an excellent energy resolution performance in high-rate applications with fast peaking time and best-in-class pile-up rejection. DANTE boxes (in 1, 8 channel or up to 32 channel custom versions, see Fig.~\ref{fig:DANTEScheme}, left) are based on the DANTE board (schematically shown in Fig.~\ref{fig:DANTEScheme}, right), connected in a daisy-chain fashion. DANTE is controlled with a C++ based library, which also offers Python and LabView compatibility.

DANTE DPP can be operated with two separated firmwares, with a different optimized feature:
\begin{itemize}
 \item \textbf{Low-Energy optimized (LE) firmware}: based on standard trapezoidal filtering, with a fixed peaking time (PT) and flat-top. This firmware is compatible with pulsed reset and continuous reset strategies. The energy resolution mainly depends on PT value and is relatively stable with ICR. In addition to that, this firmware includes a less sensitive energy threshold, optimized for low energy X-ray detection, down to Beryllium fluorescence line~\cite{Bombelli2019TowardsOX}.
 The evaluated firmware version in this work is the 4.1.1.LE.
 \item \textbf{High-Rate optimized (HR) firmware}: based on variable energy filter PT and optimized for best compromise between dead time and energy resolution. This firmware is only compatible with pulsed reset strategy. The algorithm dynamically selects the optimum PT for each detected photon between a minimum value, which defines the system dead time, and a maximum PT, which defines the best energy resolution achievable by the system. Since the high-rate firmware is based on variable PT, the energy resolution is not constant anymore with the ICR: it will be optimum and the same as for LE firmware at low ICR (1-10~kcps) but it will be progressively degrade at higher ICR values. The evaluated firmware version in this work is the 4.1.0.HR.
\end{itemize}

\section{Performance measurements with an X-ray generator source}
\label{sec:labmes}
DANTE DPP performances have been tested with an X-ray generator source and with two different Silicon Drift Detectors (SDDs): a single-element SDD (VORTEX-1EM, model HHS5508-VTX-EM500-UHV) and a four-element SDD (VORTEX-4EM, model 1303-VTX-ME4). These detectors are two good examples of all SDD detectors in operation at SOLEIL synchrotron: the first one of the previous generation before the appearance of CUBE preamplifier in 2013~\cite{6154396} and the second one of the most recent generation. Detector specifications are shown in Table~\ref{tab:SDDFeatures}. For both detectors, silicon sensors are located inside a metallic tube under vacuum and are operated at few tens of degrees Celsius below zero by a Peltier thermoelectric cooler. The tube entrance is equipped with a 8~$\mu$m-thick Beryllium window.

\begin{table}[htb!]
\centering
\begin{tabular}{l|cccc}
\hline
Detector type & VORTEX-1EM & VORTEX-4EM & BRUKER & RAYSPEC\\
\hline
Experiment & \multicolumn{2}{c}{X-ray generator source} & LUCIA & PUMA\\
\hline
Number of sensors & 1 & 4 & 1 & 1\\
Sensor thickness & 450~$\mu$m & 1~mm & 450~$\mu$m & 450~$\mu$m\\
Collimated area  & 30~mm$^2$ & 4 $\times$ 50~mm$^2$ & 60~mm$^2$ & 80~mm$^2$\\
Entrance window & Be & Be & AP3.3 & Be\\
\hline
Preamplifier type & JFET & CUBE & CUBE & CUBE\\
Preamplifier gain & 1.76~mV/keV & 1.72~mV/keV & 5.02~mV/keV & 2.42~mV/keV\\
Voltage swing & $\pm$2~V & -1.0~V/+1.5~V & $\pm$3.7~V & +1.0~V/+3.5~V\\
Reset period & 1000~ms & 28~ms & 150~ms & 95~ms\\
Reset time & 600~ns & 600~ns & 400~ns & 360~ns\\
Pulse risetime & 30.0 $\pm$ 5.4~ns & 33.3 $\pm$ 8.6~ns & 59~ns $\pm$ 17~ns & 136 $\pm$ 60~ns\\
\hline
Optimum energy & \multirow{2}{*}{136.1~$\pm$~0.2~eV} & \multirow{2}{*}{136.0~$\pm$~0.3~eV} & \multirow{2}{*}{131.3~$\pm$~0.3~eV} & \multirow{2}{*}{130.4$\pm$~0.2~eV}\\
resolution at 5.9~keV & & & & \\
Optimum PT & 8.0~$\mu$s & 1.0~$\mu$s & 2.0~$\mu$s & 2.0~$\mu$s\\
\hline
\end{tabular}
\caption{Specifications of Silicon Drift Detectors tested with DANTE DPP at X-ray generator source, LUCIA and PUMA beamlines. The reset period has been measured in absence of X-rays. The optimum energy resolution at 5.9~keV has been measured with a XIA-DXP-XMAP DPP~\cite{XIAXMAP,Hubbard1996}, and using a $^{55}$Fe source.}
\label{tab:SDDFeatures}
\end{table}

The main difference between the two detectors is the front-end electronics: the VORTEX-1EM is equipped with a preamplifier with a standard JFET transistor;
the VORTEX-4EM uses CUBE preamplifier, which is fully implemented in a CMOS process and is less sensible to pickup noise and interferences. This difference in front-end electronics translates to detector performance in terms of an optimum energy resolution at a short PT value for the SDD equipped with CUBE: 1.0~$\mu$s for VORTEX-4EM vs 8.0~$\mu$s for VORTEX-1EM. The fact of keeping good energy resolution (i.e., satisfactory spectroscopy performance) at short PT values allows the use of SDD detectors equipped with CUBE preamplifier at higher ICR, keeping the dead time in a reasonable level and maximimizing the OCR~\cite{6551138}.

\subsection{Experimental setup and procedure}
\label{sec:labexp}
\begin{figure}[htb!]
\centering
\includegraphics[width=71mm]{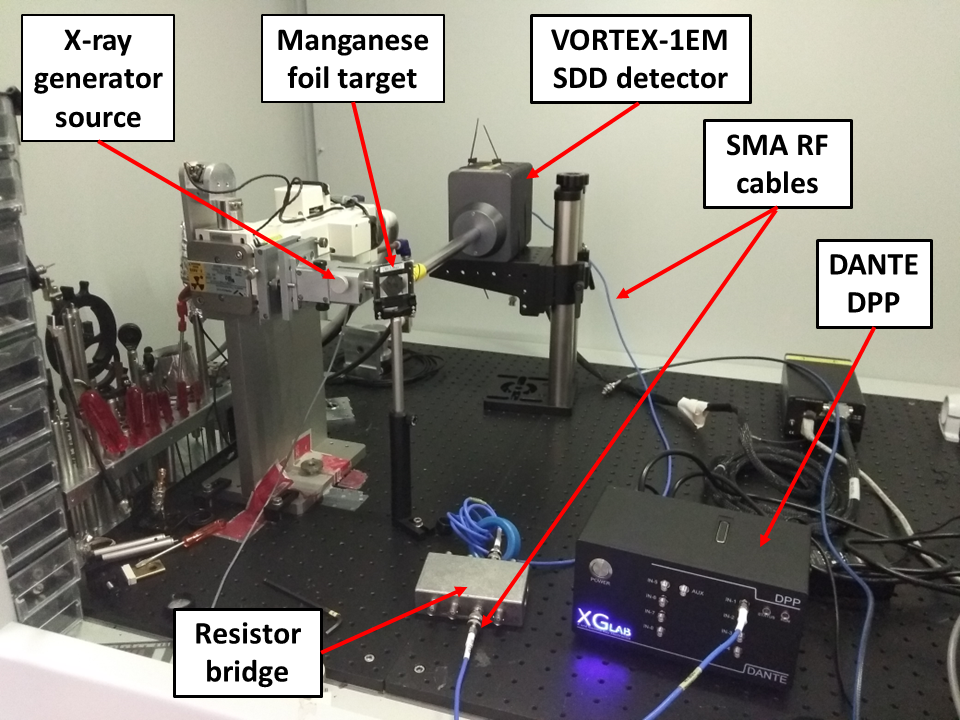}
\includegraphics[width=77mm]{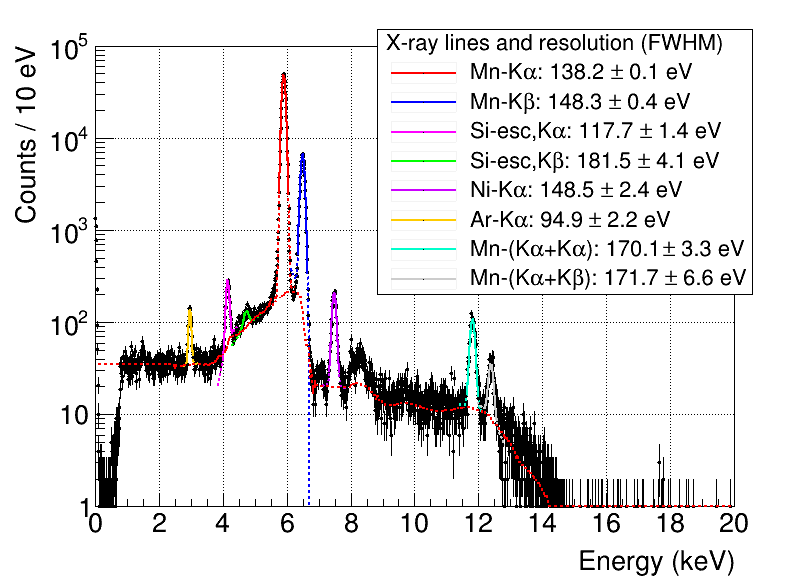}
\caption{Left: View of the experimental setup during the test of VORTEX-1EM SDD. Right: Energy spectrum of a metallic manganese foil acquired by VORTEX-1EM SDD detector read by DANTE DPP (PT of 6016 ns) at an X-ray generator setting of 15~kV and 4~mA (ICR of 78.5~kcps). Different X-ray lines (fluorescence, escape and pile-up) have been identified and adjusted by a Gaussian function. Energy resolution (FHWM) is indicated for each line.}
\label{fig:LabEnerSpec}
\end{figure}

Each SDD detector has been installed inside an X-ray generator hutch (see Figure~\ref{fig:LabEnerSpec}, left), with its window perpendicular to the beam. A manganese foil target (4~$\mu$m thick foil deposited on a polyethylene base), situated in the beam at 45~degrees, is used to generated fluorescence X-rays. In the case of VORTEX-4EM SDD, the first element (top-left one) has been aligned to the target to maximize ICR and performance results are only given for this element. Detector preamplifier outputs have been connected to the input channels of a 8-channel DANTE DPP by SMA cables and, in the case of VORTEX-1EM SDD, a resistor bridge to adapt the detector output swing (4~Vpp) to DPP one (3~Vpp). An energy spectrum of 30~seconds acquisition time has been recorded for each X-ray generator settings (current values between 0 and 40~mA and for 15 and 50~kV voltage values), equivalent to an ICR between 10 and 740~(430)~kcps and between 67 and 5120~(3080)~kcps for VORTEX-1EM (VORTEX-4EM) SDD, respectively.

Before tests, DANTE digital gain has been adjusted using the LE firmware at a PT of 1024~ns and a flat top of 192~ns for VORTEX-1EM and 128~ns for VORTEX-4EM, in such a way that the manganese fluorescence X-ray line (at 5.9 keV) is situated at channel of 686, equivalent to a spectrum binning of 10~eV and a default offset of 96 channels. The specific values of the digital gain are 4.562 for VORTEX-1EM SDD and 2.978 for VORTEX-4EM SDD. During the tests, the following DANTE DPP settings have been varied:
\begin{itemize}
 \item \textbf{LE firmware}: PT varied between 96 and 9984~ns for VORTEX-1EM SDD and between 96 and 4096~ns for VORTEX-4EM SDD.
 \item \textbf{HR firmware}: The minimum PT varied between 32 and 192~ns, while the maximum PT was set to 2048~ns for VORTEX-1EM SDD and 1024~ns for VORTEX-4EM SDD. For the second SDD, this value corresponds to the optimum PT, while for the first SDD, it is the maximum PT value that can be set by the electronics.
\end{itemize}

Each energy spectrum (an example is shown in Figure~\ref{fig:LabEnerSpec}, right) is composed of two intense manganese fluorescence X-ray lines (K$_\alpha$-line at 5.9~keV and K$_\beta$-line at 6.4~keV), a continuous Compton/Rayleigh scattering background generated by the X-ray generator beam, two escape peaks situated at a lower energy from main lines (1.74~keV less, at 4.16 and 4.66~keV) and generated by the silicon sensor, argon fluorescence X-ray line from air (K$_\alpha$-line at 2.96~keV), nickel fluorescence X-ray line from sample support (at 7.472~keV) and two pile-up lines on the right side of the spectrum, generated by random detection coincidences (called pile-up) of manganese lines (K$_\alpha$-K$_\alpha$ at 11.8~keV and K$_\alpha$-K$_\beta$ at 12.3~keV). In the offline analysis of each spectrum based on ROOT\cite{Brun:1997pa}, the continuous background has been fitted and substracted, and X-ray lines have been identified and adjusted by a Gaussian function to estimate the peak position (mean value) and energy resolution (FWHM value). For the manganese K$_\alpha$-line, the statistical error of energy resolution is less than 0.1\% and is negligible for peak position.

The performance of the combination of the two DANTE firmwares and the two detectors has been characterized in terms of:
\begin{enumerate}
    \item Energy resolution at 5.9~keV at ICR values of 10~kcps and 1~Mcps.
    \item The peak stability between 10~kcps and 1~Mcps.
    \item The measurement time ($\tau_1$), defined as the minimum interval time between two successive impulses that is acceptable by the DPP for a correct reconstruction of energy values~\cite{Bordessoule2019}. This parameter is calculated from the dependence of OCR and ICR, using the expression:
\begin{equation}
 OCR = ICR \times \exp(-\tau_1 \times ICR)
\label{eq:tau1}
\end{equation}
    The dead time at 1~Mcps, provided in DPP notices, is calculated from measurement time.
    \item The time resolution ($\tau_2$, or pile-up rejection power), defined as the minimum interval time between two successive impulses for which the DPP can separate them~\cite{Bordessoule2019}. This parameter is estimated from the dependence of the pile-up ratio of intensities (i.e., the ratio of intensities of the K$_\alpha$-K$_\alpha$ pile-up line ($Int(2)$) and the $K_\alpha$ line ($Int(1)$) with the ICR, using the expression
\begin{equation}
 \frac{Int(2)}{Int(1)} = ICR \times \tau_2 / 2
 \label{eq:tau2}
\end{equation}
    The pile-up intensity at 1~Mcps, provided in DPP notices, is calculated from time resolution.
\end{enumerate}

\subsection{Performance results of Low Energy optimized (LE) firmware}
\label{sec:labreslef}
\begin{figure}[htb!]
\centering
\includegraphics[width=75mm]{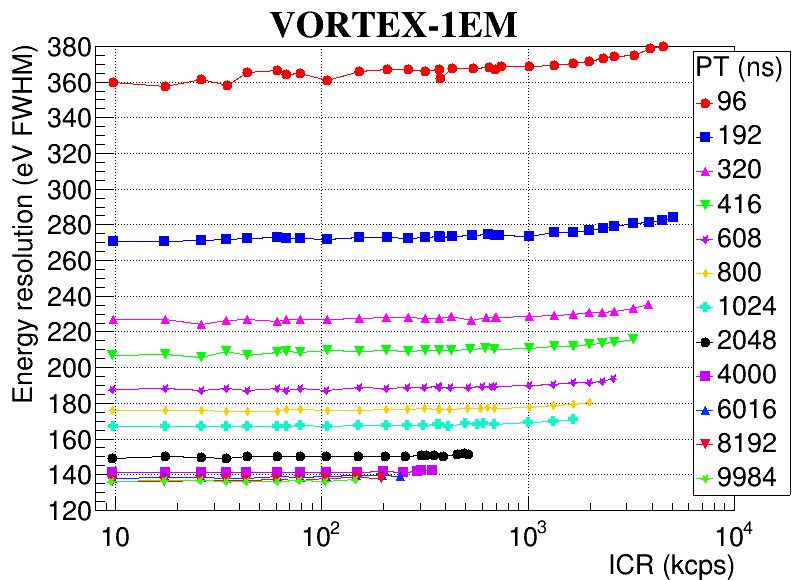}
\includegraphics[width=75mm]{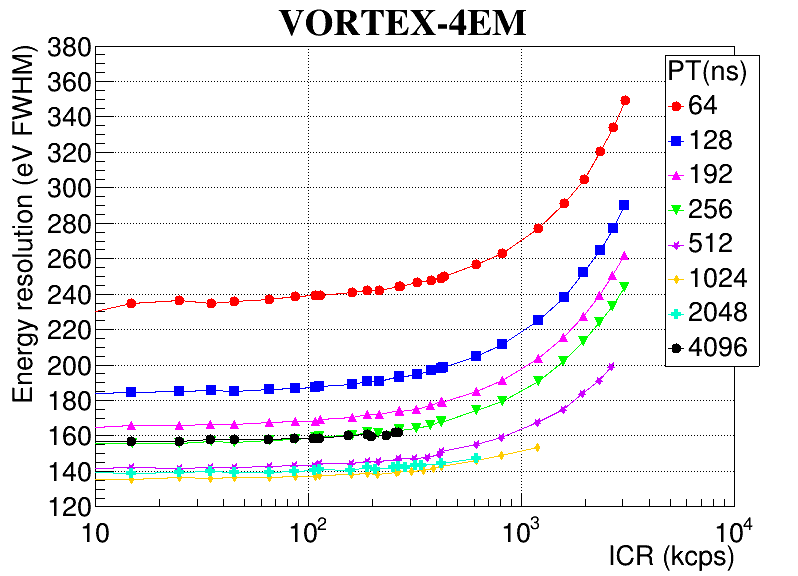}
\includegraphics[width=75mm]{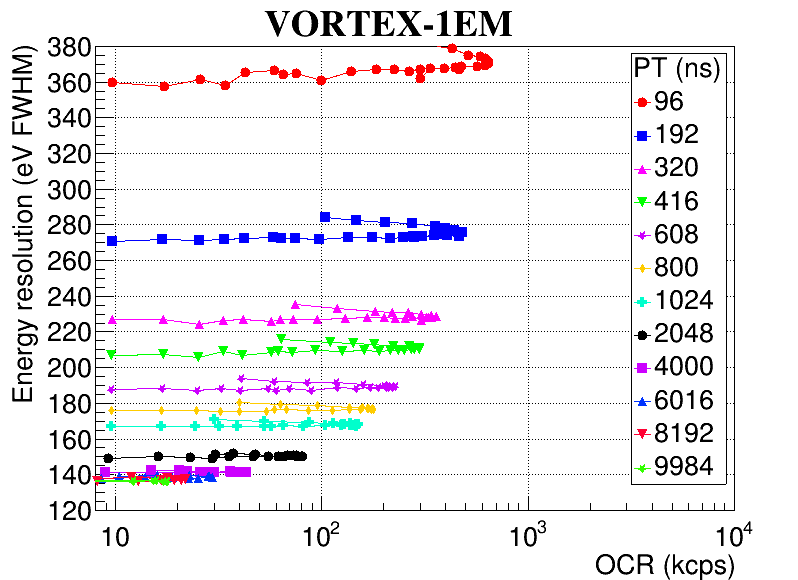}
\includegraphics[width=75mm]{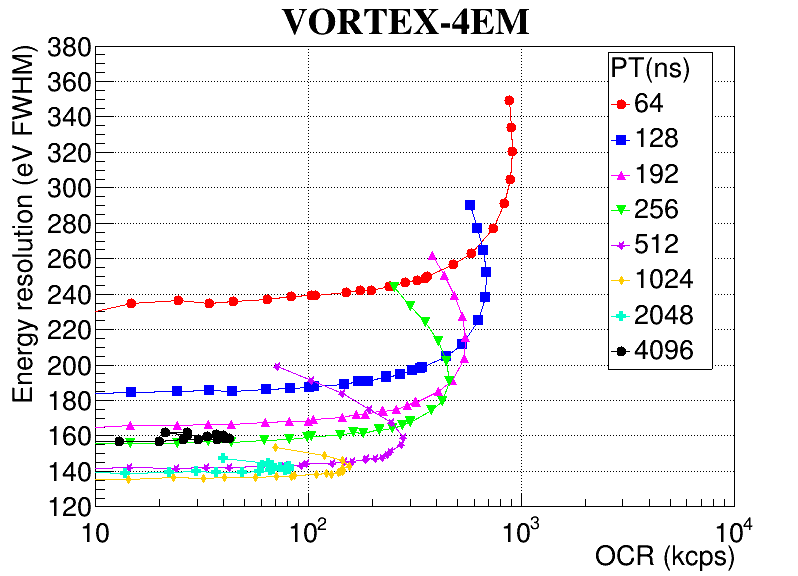}
\caption{Dependence of the energy resolution (FHWM) at 5.9~keV
with ICR (top) and OCR (bottom),
measured by VORTEX-1EM (left) and VORTEX-4EM SDDs (right), read by DANTE DPP for different PT values.}
\label{fig:LabEnerResvsICROCR}
\end{figure}

The dependence of the energy resolution (FWHM) with ICR and OCR for different PT values is shown in Figure~\ref{fig:LabEnerResvsICROCR}. At a given PT value, the energy resolution is constant for VORTEX-1EM SDD (less than 1\% deviation from the mean value) for ICR values between 10 and 1000~kcps and it slightly degrades at higher ICR values, up to a 4\% from 10~kcps to 5~Mcps. For the VORTEX-4EM SDD, energy resolution is stable in a shorter ICR range, between 10 and 100~kcps, and it degrades up to 17\% from 10~kcps to 1~Mcps.

\begin{figure}[htb!]
\centering
\includegraphics[width=75mm]{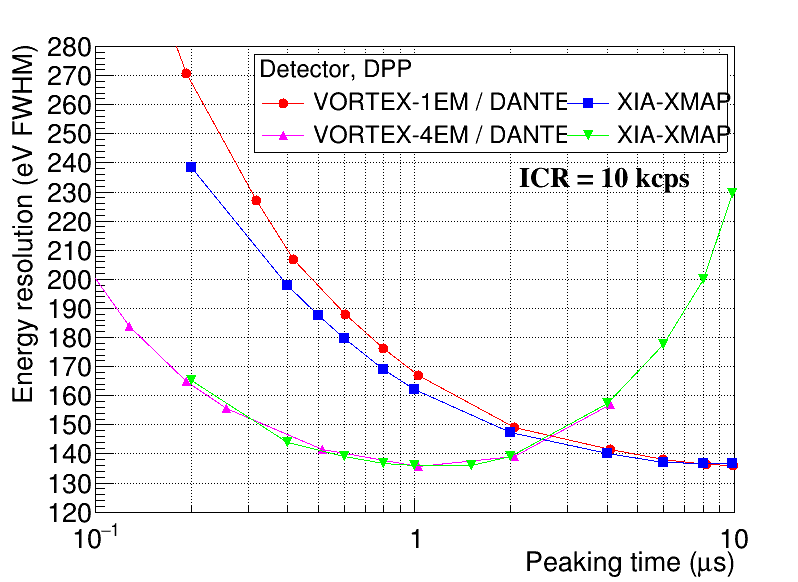}
\includegraphics[width=75mm]{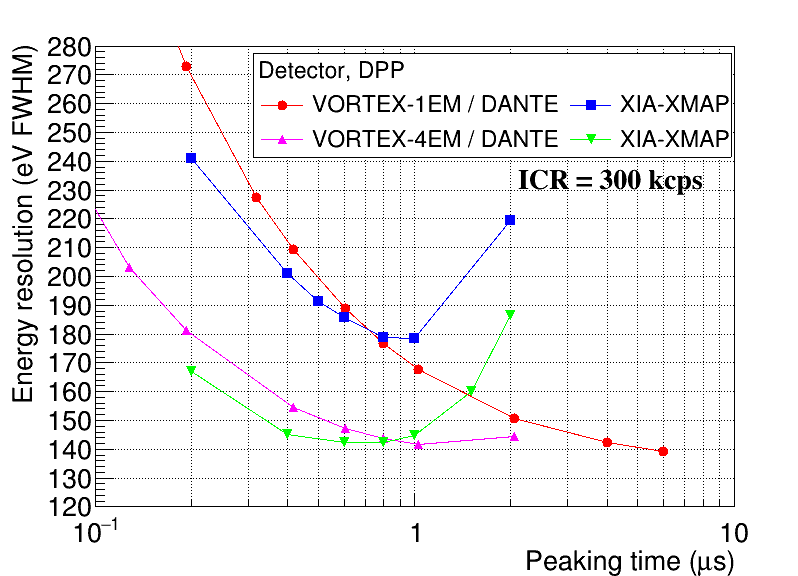}
\caption{Dependence of the energy resolution (FHWM) at 5.9~keV with PT at an ICR of 10 (left) and 300~kcps (right) for VORTEX-1EM and VORTEX-4EM SDDs, and DANTE and XIA-XMAP DPPs.}
\label{fig:LabPeakPosvsPT}
\end{figure}

The specific value of energy resolution depends on the PT, as shown in Figure~\ref{fig:LabPeakPosvsPT} for an ICR of 10~kcps and 300~kcps. For the VORTEX-1EM SDD, the best value at 10~kcps (136.4~$\pm$~0.2~eV) has been measured at a PT of 8192~ns and is compatible with the reference value (136.1~$\pm$~0.2~eV), measured with XIA-DXP-XMAP DPP. For the VORTEX-4EM SDD, the best value (135.7~$\pm$~0.2~eV) has been measured at a PT of 1024~ns and is also compatible with the reference value (136.0~$\pm$~0.3~eV). Energy resolution measurements by DANTE DPP are compatible with those made by XIA-DXP-XMAP DPP at 10~kcps, except for VORTEX-1EM SDD and PT values shorter than 1000~ns, for which the result difference varies between 5~eV at 1024~ns and 30~eV at 200~ns. At a higher ICR of 300~kcps, DANTE keeps an optimum energy resolution at PT longer than 1000~ns, while XIA-XMAP becomes saturated. In contrast, XIA-XMAP provides slightly better energy resolution at PT shorter than 600~ns, the difference is around 12~eV for a PT of 200 ns.

\begin{figure}[htb!]
\centering
\includegraphics[width=75mm]{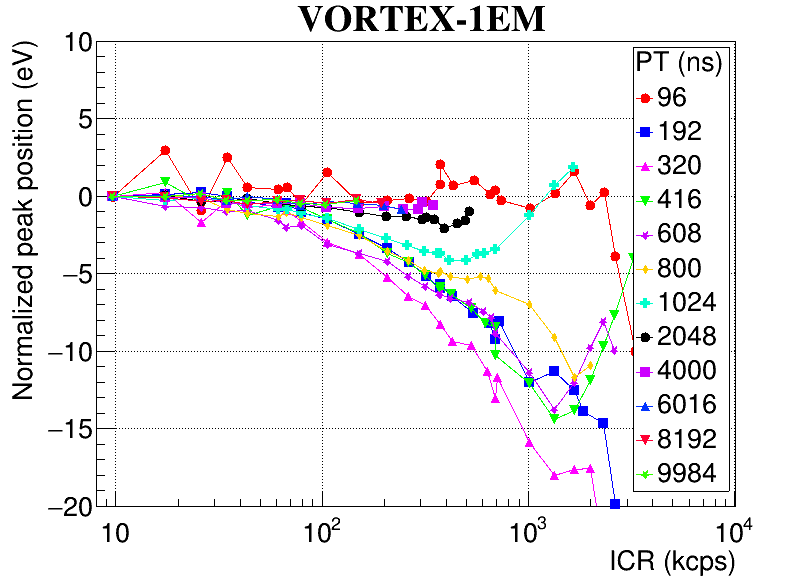}
\includegraphics[width=75mm]{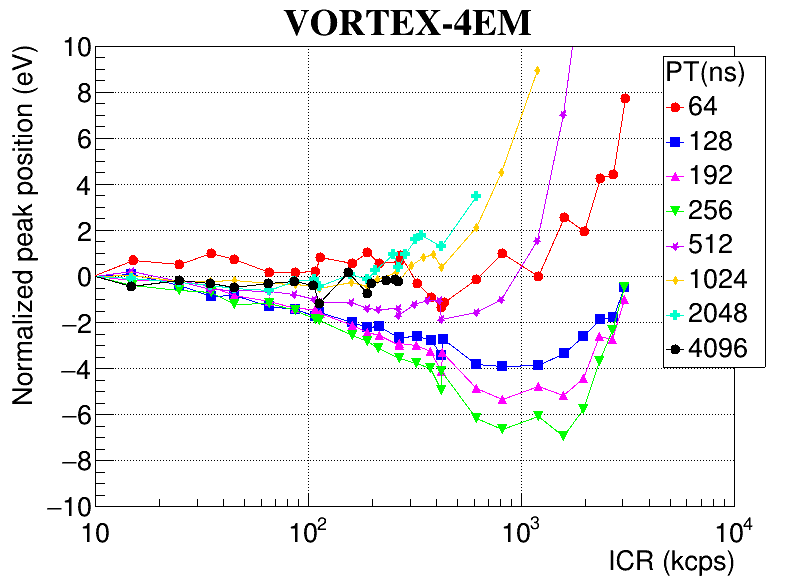}
\includegraphics[width=75mm]{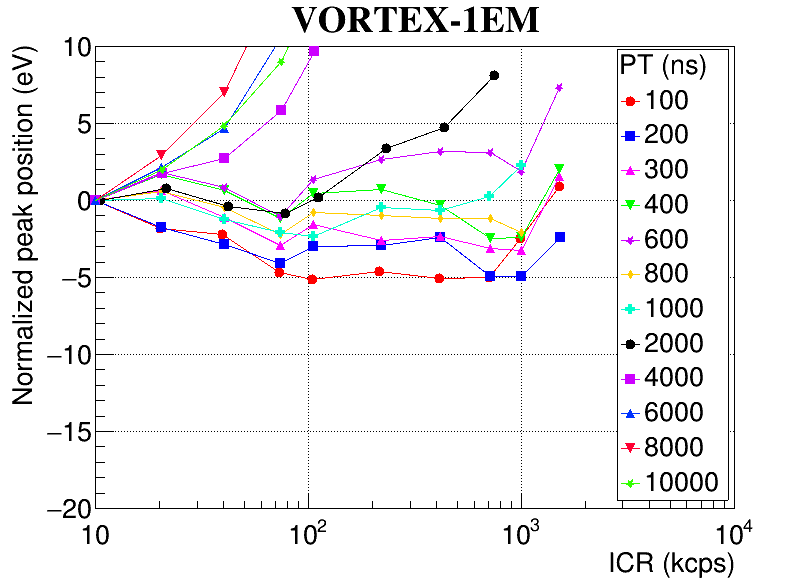}
\includegraphics[width=75mm]{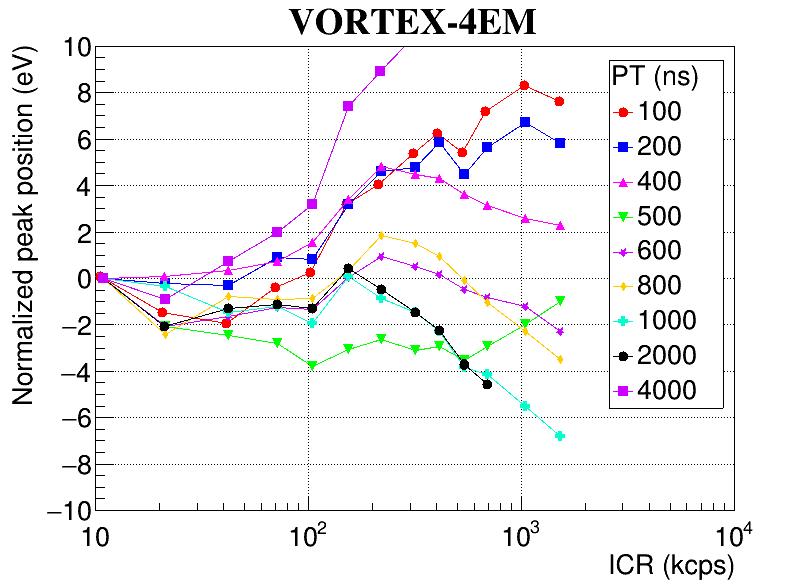}
\caption{Dependence of the normalized peak position at 5.9~keV with ICR, measured by VORTEX-1EM (left) and VORTEX-4EM SDD (right), read by DANTE (top) and XIA-XMAP DPP (bottom) for different PT values.}
\label{fig:LabPeakPosvsICR}
\end{figure}

The stability of the manganese K$_\alpha$-line with the photon flux has been expressed in terms of the normalized peak position, defined as the difference between the peak position at a given ICR and the estimated one at 10~kcps. The dependence of the normalized position with ICR and for different PT values is shown in Figure~\ref{fig:LabPeakPosvsICR} (top part). For VORTEX-1EM SDD and all PT values, the normalized peak position becomes negative, i.e., the peak shifts to lower energy values if ICR is varied from 10 to 1000~kcps. Peak shift absolute values in this ICR range are smaller than 16~eV (measured for a PT of 320~ns), i.e., a peak variation of less than 0.3\%, well within specifications for synchrotron applications (less than 1\%). In the case of VORTEX-4EM SDD, the normalized peak position becomes negative between 10 and 1000~kcps, and then positive at higher ICR values. Absolute shifts values are smaller for this detector than those of VORTEX-1EM SDD, up to 7~eV for the same ICR range (measured for a PT of 1024~ns), i.e., a variation of less than 0.1\%, which is considered negligible. The results for both detectors are also compatible with measured peak shift values by XIA-DXP-XMAP (bottom part): up to 5~eV (measured for 200~ns) for VORTEX-1EM SDD and 8~eV (measured for 100~ns) for VORTEX-4EM SDD.

\begin{figure}[htb!]
\centering
\includegraphics[width=75mm]{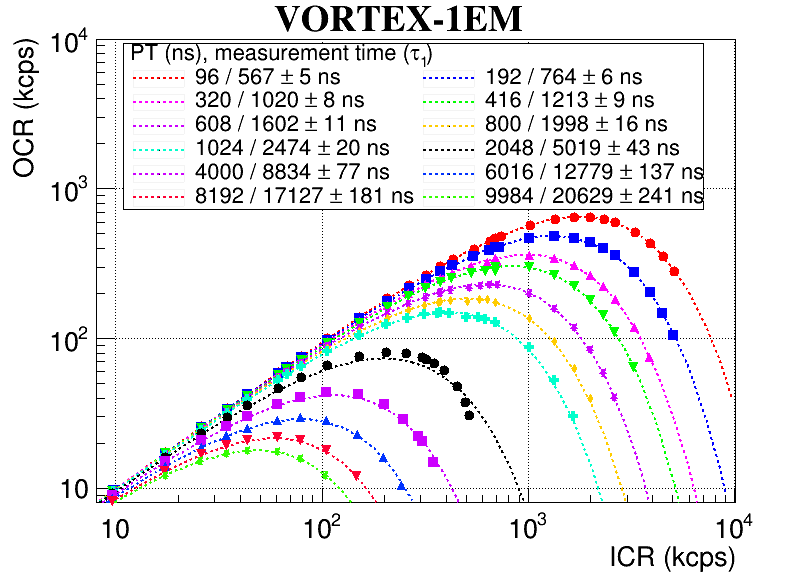}
\includegraphics[width=75mm]{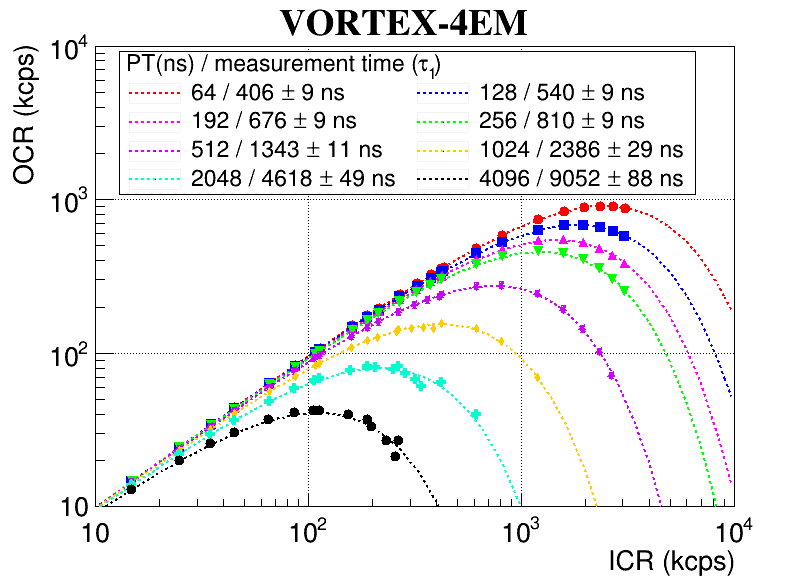}
\caption{Dependence of the OCR with ICR, measured by VORTEX-1EM (left) and VORTEX-4EM SDD (right), read by DANTE DPP for different PT values. Each dashed line is the fit result of expressions~\ref{eq:tau1} to experimental data. Measurement time values are shown in each plot legend.}
\label{fig:LabOCRvsICR}
\end{figure}

The linearity of OCR vs ICR for different PT values is shown in Figure~\ref{fig:LabOCRvsICR}. Measurement time values for VORTEX-1EM SDD detector vary from 570 to 20530~ns for PT values between 96 and 9984~ns, while for VORTEX-4EM SDD they range from 410 to 9050~ns. The fact that VORTEX-4EM SDD detector shows shorter measurement times than VORTEX-1EM SDD is explained by the better performance of its preamplifier at short PT values, as discussed in Sec~\ref{sec:labmes}. The results for both detectors are also compatible with measured values by a XIA-DXP-XMAP DPP.

\begin{figure}[htb!]
\centering
\includegraphics[width=75mm]{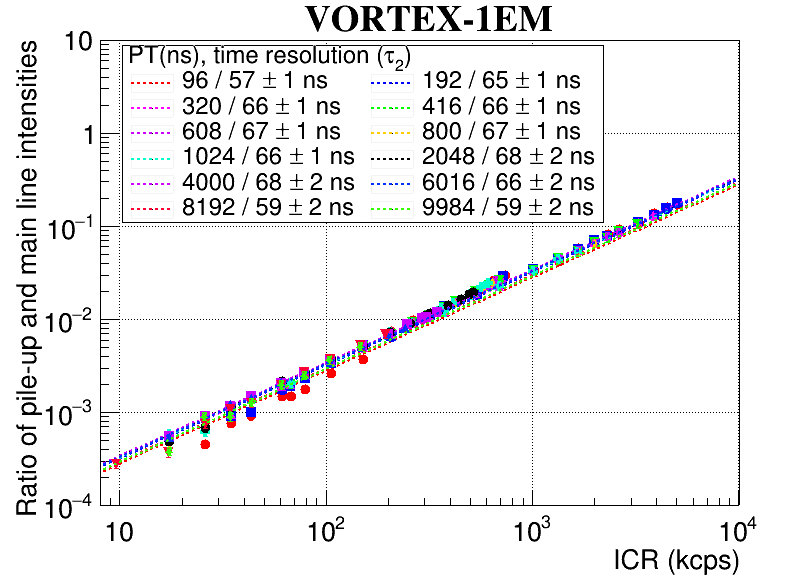}
\includegraphics[width=75mm]{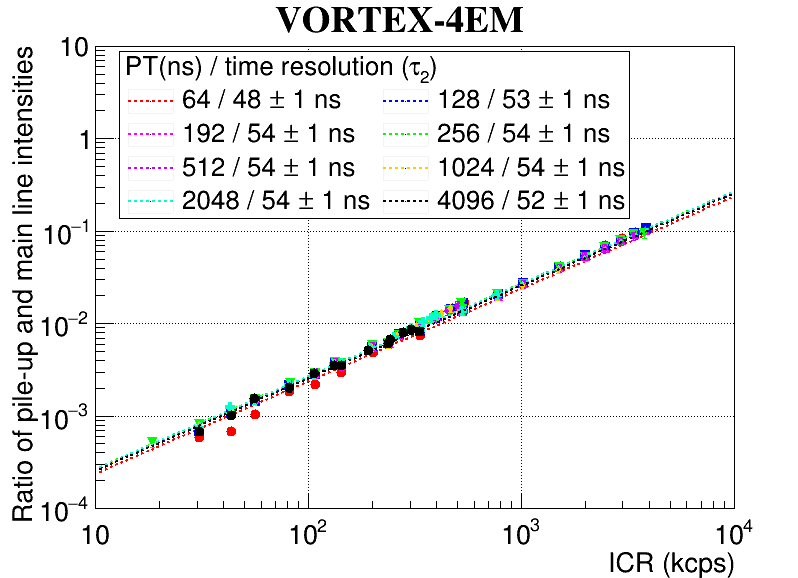}
\caption{Dependence of the pile-up ratio of intensities with ICR, measured by VORTEX-1EM (left) and VORTEX-4EM SDD (right), read by DANTE DPP for different PT values. Each dashed line is the fit result of expression~\ref{eq:tau2} to experimental data. Time resolution values are shown in each plot legend.}
\label{fig:LabIntvsICR}
\end{figure}

The dependence of the pile-up ratio of intensities and ICR for different PT values is shown in Figure~\ref{fig:LabIntvsICR}. The estimated time resolution for VORTEX-1EM SDD is 67~$\pm$~2~ns for all PT values, except for 96, 8192 and 9984~ns where the estimation is probably underestimated (57-59~ns). For the PT of 96~ns, pile-up line area is underestimated because of a degraded energy resolution. For the other two PT values, a systematic fit error caused by the limited number of experimental points can explain the short time. The estimated time resolution for VORTEX-4EM SDD is 54~$\pm$~1~ns, except for a PT of 96~ns, where the estimation is also underestimated (48~ns) by a degraded energy resolution. For both SDD detectors, estimated values for time resolution are much better than those measured by XIA-DXP-XMAP DPP (271~$\pm$~8~ns and 166~$\pm$~6~ns for VORTEX-1EM and -4EM, respectively), i.e., DANTE DPP separates much better pile-up pulses.

\subsection{Performance results of High Rate optimized (HR) firmware}
\label{sec:labreshrf}
The performance of the HR firmware of DANTE DPP has been characterized in the same terms as LE firmware, in order to confirm its higher OCR (and smaller deadtime) compared to LE firmware for the same (maximum) PT, keeping a good energy resolution, an optimal peak stability and pile-up rejection.

\begin{figure}[htb!]
\centering
\includegraphics[width=75mm]{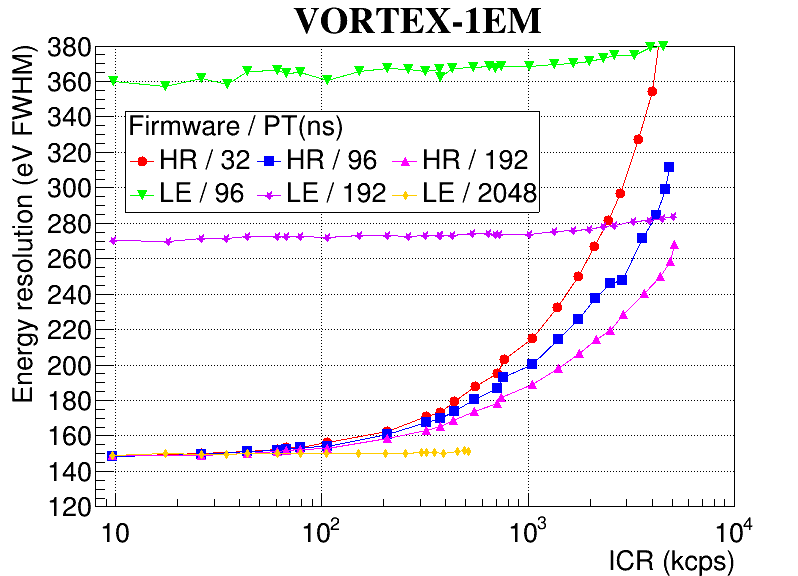}
\includegraphics[width=75mm]{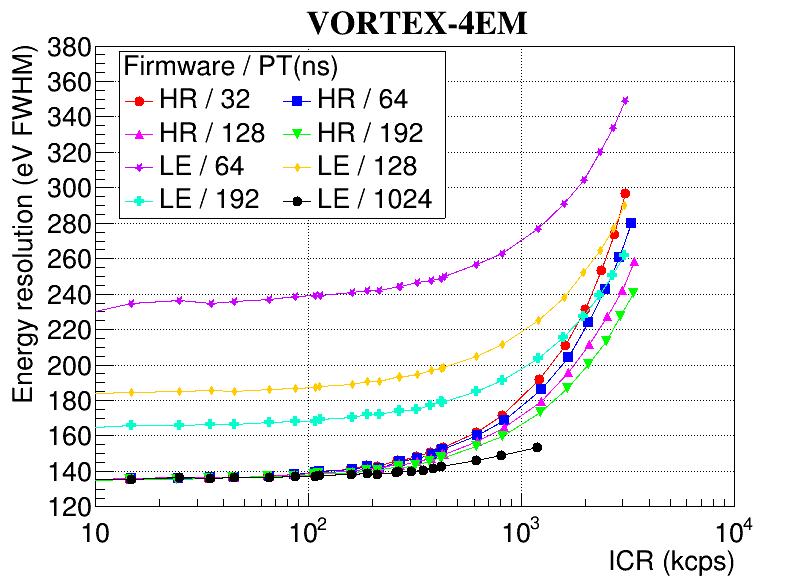}
\includegraphics[width=75mm]{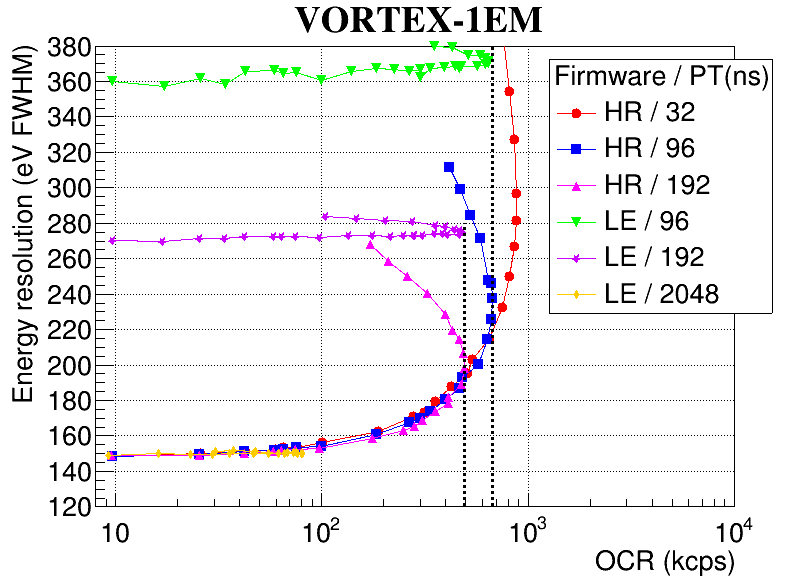}
\includegraphics[width=75mm]{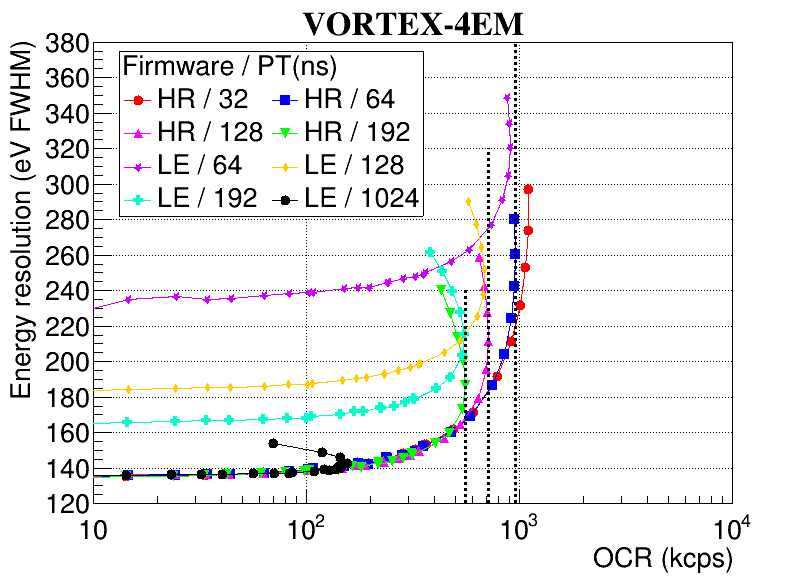}
\caption{Dependence of the energy resolution (FHWM) at 5.9~keV with ICR (top) and OCR (bottom), measured by VORTEX-1EM (left) and VORTEX-4EM SDD (right), read by DANTE DPP for HR and LE firmwares and different PT values. For OCR plots, dashed lines mark the maximum OCR measured by HR firmware at a PT of 96 and 192~ns for VORTEX-1EM (left) and at a PT of 64, 128 and 192~ns for VORTEX-4EM (right).}
\label{fig:LabEnerResvsICROCRFast}
\end{figure}

The dependence of the energy resolution at 5.9~keV (FWHM) with ICR and OCR and for the different minimum PT values is shown in Figure~\ref{fig:LabEnerResvsICROCRFast}. For HR firmware and all PT values, the energy resolution at 10~kcps is the same as the one measured for the maximum PT and LE firmware, i.e., 148.3~$\pm$~0.3~eV for VORTEX-1EM SDD. The energy resolution is not optimal because the maximum PT (2048~ns) is not the optimum PT (8000 ns). For the VORTEX-4EM SDD, the energy resolution at 10~kcps is 135.1~$\pm$~0.2~eV. In this case, the resolution is optimal because the maximum PT (1024~ns) is the optimum. At ICR values higher than 100~kcps, energy resolution of HR firmware and for two detectors progressively degrades with ICR, as expected. However, the energy resolution is better than that of LE firmware up to the maximum OCR. This fact is illustrated by the dashed lines in the bottom plots of Figure~\ref{fig:LabEnerResvsICROCRFast}, which mark the maximum OCR measured by HR firmware at common PT values.

\begin{figure}[htb!]
\centering
\includegraphics[width=75mm]{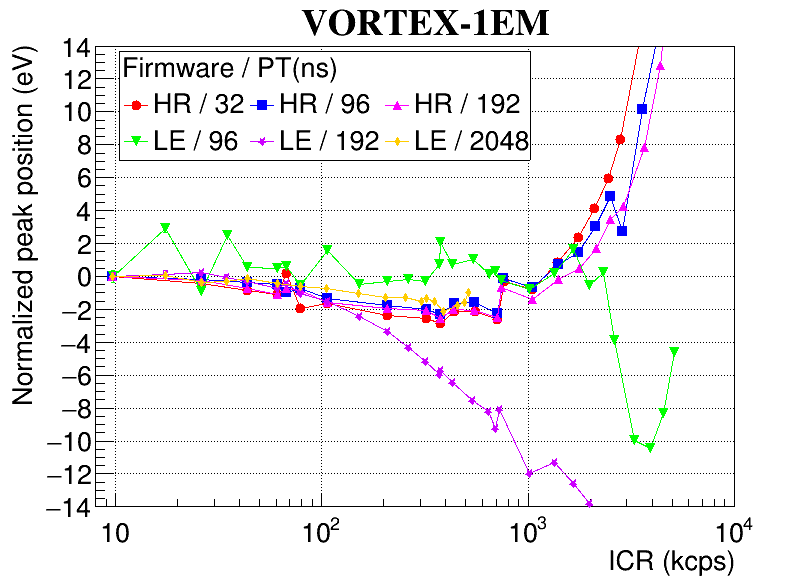}
\includegraphics[width=75mm]{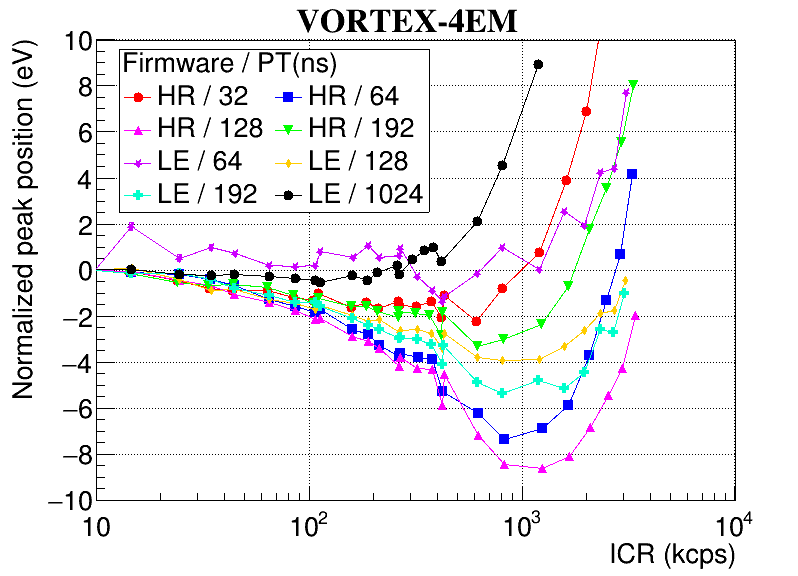}
\caption{Dependence of the normalized peak position at 5.9~keV with ICR, measured by VORTEX-1EM (left) and VORTEX-4EM SDD (right), read by DANTE DPP for HR and LE firmwares and different PT values.}
\label{fig:LabPeakPosvsICRFast}
\end{figure}

The dependence of the normalized position with ICR and for different minimum PT values is shown in Figure~\ref{fig:LabPeakPosvsICRFast}. For HR firmware and both detectors, the normalized peak position becomes negative between 10 and 1000~kcps, and then positive at higher ICR values. Peak shift absolute values in the ICR range of 10 to 1000~kcps are smaller than 2.9~eV for VORTEX-1EM SDD (measured for a minimum PT of 32~ns) and 8.5~eV for VORTEX-4EM SDD (for 128~ns), i.e., a variation of less than 0.05\% and 0.15\%, respectively. For the two detectors, the peak shift has not enlarged compared to LE firmware and results are well within specifications for synchrotron applications (less than 1\%).

\begin{figure}[htb!]
\centering
\includegraphics[width=75mm]{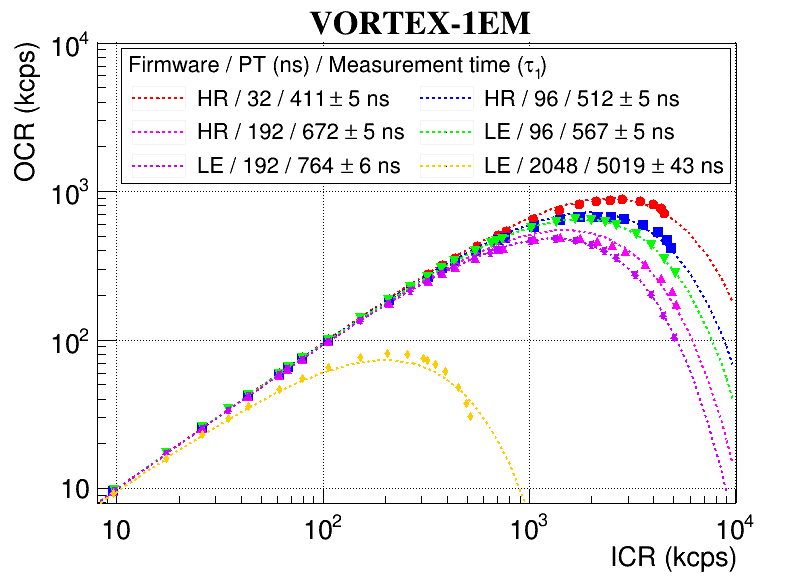}
\includegraphics[width=75mm]{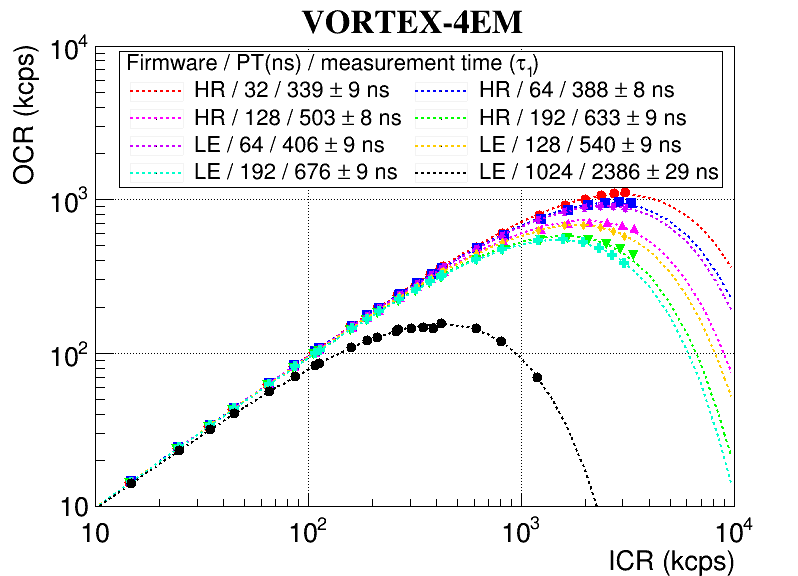}
\caption{The linearity of OCR vs ICR, measured by VORTEX-1EM (left) and VORTEX-4EM SDD (right), read by DANTE DPP for HR and LE firmwares and different PT values. Each dashed line is the fit result of expressions \ref{eq:tau1} to experimental data. Measurement time values are shown in each plot legend.}
\label{fig:LabOCRvsICRFast}
\end{figure}

The higher OCR of HR firmware compared to LE firmware is confirmed by the linearity of OCR vs ICR for different minimum PT values, shown in Figure~\ref{fig:LabOCRvsICRFast}. For the VORTEX-1EM SDD detector and the HR firmware, the measurement time value is almost compatible with the estimated value for the same PT and LE firmware, i.e., the OCR is defined by the minimum PT and not the maximum PT, as expected. In the case of the VORTEX-4EM SDD, the measurement time values is a slightly better (10\%) for HR firmware than for LE one.

\begin{figure}[htb!]
\centering
\includegraphics[width=75mm]{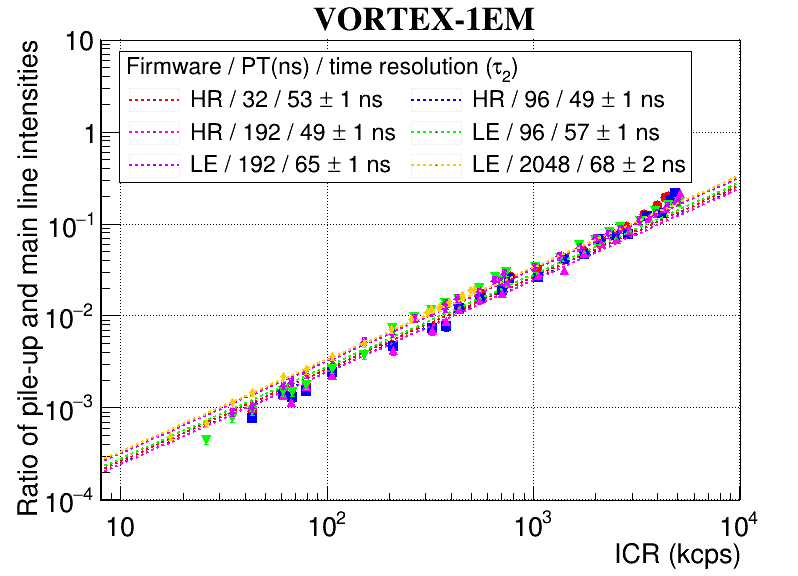}
\includegraphics[width=75mm]{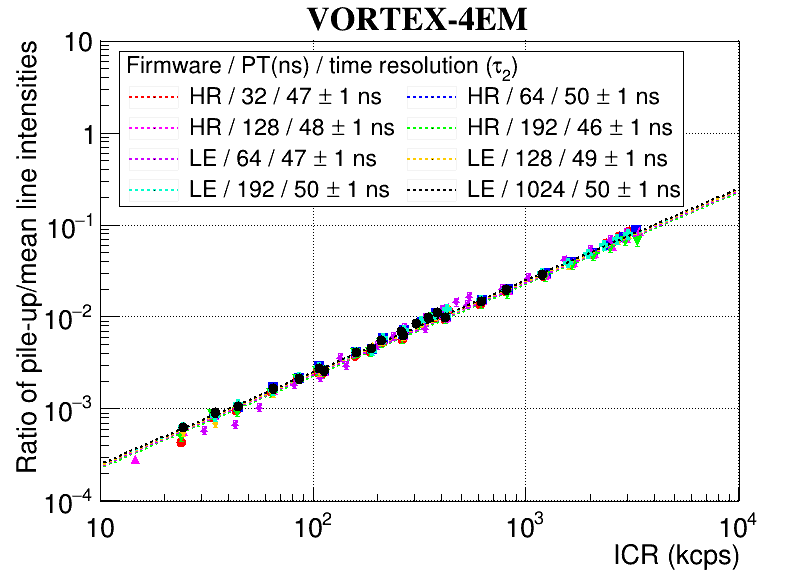}
\caption{Dependence of the pile-up ratio of intensities with ICR, measured by VORTEX-1EM (left) and VORTEX-4EM SDD (right), read by DANTE DPP for HR and LE firmwares and different PT values. Each dashed line is the fit result of expression \ref{eq:tau2} to experimental data. Time resolution values are shown in each plot legend.}
\label{fig:LabIntvsICRFast}
\end{figure}

Finally, the dependence of the pile-up ratio of intensities and ICR for different minimum PT values is shown in Figure~\ref{fig:LabIntvsICRFast}. The measured time resolution for HR firmware is $\approx$50~ns for the two detectors. In the case of VORTEX-1EM SDD, the time resolution value is better than that of LE firmware (67~$\pm$~2~ns), while for VORTEX-4EM SDD the two values are the same. In conclusion, DANTE shows the same or slightly better pile-up rejection power for HR firmware compared to LE firmware.

\subsection{Comparison with other recent DPP}
\label{sec:compdpp}
Two recent DPPs, XIA-FalconX and Xspress3, have been tested with VORTEX-1EM and VORTEX-4EM SDDs in the same experimental setup in order to compare the performance of the three DPPs. A calibration file was optimized for each combination of DPP and SDD detector following user manuals\footnote{In the case of XIA-FalconX, the pulse characterization was made at an analog gain of 2.0 and a very short decay time, and a low rate filter was used.}. The results are shown in Table~\ref{tab:DANTEComp}. DANTE with HR firmware shows an optimum energy resolution at 10~kcps but the resolution degrades with ICR. This degradation is similar to that of other DPPs for VORTEX-1EM SDD and slightly faster for VORTEX-4EM SDD. Peak shifts are smaller than those of the two other DPPs. XIA-FalconX and specially the Xpsress3 show a shorter measurement time and lower dead time at 1~Mcps than DANTE DPP, i.e., they provide a higher OCR. Finally, DANTE shows a shorter time resolution and reduced pile-up intensity at 1~Mcps than that of the other DPPs, except for VORTEX-4EM SDD, where DANTE and XIA-FalconX show similar time resolution values.

\begin{table}[htb!]
\centering
\begin{tabular}{cc}
\rotatebox[origin=c]{90}{VORTEX-1EM} &
\begin{tabular}{l|ccc}
\hline
Parameter & DANTE & XIA-FalconX & Xspress3\\
\hline
Energy resolution at 10~kcps & 148.2$\pm$0.3 eV & 178.6$\pm$0.3 eV & 163.3$\pm$0.2 eV\\
Energy resolution at 1~Mcps & 214.8$\pm$0.1 eV & 245.2$\pm$0.1 eV & 213.1$\pm$0.1 eV\\
\hline
\multirow{2}{*}{Peak shift 10 kcps - 1 Mcps} & -0.8$\pm$0.1 eV & -5.7$\pm$0.1~eV & -10.7$\pm$0.1 eV\\
 & 0.045\% & -0.097\% & -0.181\% \\
\hline
Measurement time & 411$\pm$5~ns & 199$\pm$4~ns & 77$\pm$12~ns\\
Dead time at 1 Mcps & 33.7$\pm$0.3\% & 18.0$\pm$0.3\% & 7.4$\pm$1.1\%\\
OCR at 1~Mcps & 663 kcps & 820 kcps & 926 kcps\\
\hline
Time resolution & 53$\pm$1 ns & 72$\pm$1 ns & 99$\pm$2 ns\\
Pile-up intensity at 1 Mcps & 2.7$\pm$0.1\% & 3.6$\pm$0.1\% & 5.0$\pm$0.1\%\\
\end{tabular}\\
\hline
\hline
\rotatebox[origin=c]{90}{VORTEX-4EM} &
\begin{tabular}{l|ccc}
Energy resolution at 10~kcps & 135.1$\pm$0.2 eV & 133.1$\pm$0.2 eV & 139.3$\pm$0.2 eV\\
Energy resolution at 1~Mcps & 181.5$\pm$0.1 eV & 153.1$\pm$0.1 eV & 162.9$\pm$0.1 eV\\
\hline
\multirow{2}{*}{Peak shift 10 kcps - 1 Mcps} & -0.1$\pm$0.1 eV & -7.4$\pm$0.1~eV & 5.2$\pm$0.1 eV\\
 & -0.002\% & -0.125\% & 0.088\% \\
\hline
Measurement time & 339$\pm$9~ns & 193$\pm$12~ns & 66$\pm$13~ns\\
Dead time at 1~Mcps & 28.8$\pm$0.6\% & 17.6$\pm$1.0\% & 6.4$\pm$1.2\%\\
OCR at 1~Mcps & 712 kcps & 824 kcps & 936 kcps\\
\hline
Time resolution & 47$\pm$1 ns & 43$\pm$1 ns & 93$\pm$2 ns\\
Pile-up intensity at 1 Mcps & 2.4$\pm$0.1\% & 2.2$\pm$0.1\% & 4.7$\pm$0.1\%\\
\hline
\end{tabular}\\
\end{tabular}
\caption{Summary of the performance results of VORTEX-1EM and VORTEX-4EM SDDs read by DANTE (with HR firmware and minimum PT of 32~ns), XIA-FalconX and Xspress3 DPPs.}
\label{tab:DANTEComp}
\end{table}

\section{XRF and XAS experiments at LUCIA beamline}
\label{sec:LUCIAint}
LUCIA beamline~\cite{FLANK2006269,Vantelon:vv5122} at SOLEIL synchrotron is a tender X-ray beamline working in the 0.8-8~keV energy range with capabilities for chemical speciation by macro or micro X-ray absorption spectroscopy ($\mu$-XAS) and  for elemental mapping by micro X-ray fluorescence ($\mu$-XRF). When micrometric, the X-ray beam is focused on the sample by two mirrors in Krikpatrick-Baez (KB) configuration, leading to an effective beam size of $2.5 \times 2.5$~$\mu$m$^2$ FWHM. The KB mirrors are retractable to allow working with a millimeter size beam (around $2.5 \times 2.5$ mm$^2$). Analyzed samples are installed on a motorized stage for alignment and surface scan situated in a vacuum chamber, operated at a vacuum level between $10^{-2}$ and $10^{-6}$~mbar. DANTE DPP was compared to actual beamline DPP, XIA-XMAP, in two different XRF and XAS experiments.

%LUCIA beamline~\cite{FLANK2006269,Vantelon:vv5122} at SOLEIL synchrotron is a soft to tender X-ray beamline (0.8-8~keV) with capabilities for chemical speciation by micro X-ray absorption spectroscopy ($\mu$-XAS) and for elemental mapping by X-ray micro-fluorescence ($\mu$-XRF). Samples are analyzed in a vacuum chamber at a vacuum level around $10^{-4}$~mbar to have access to soft X-ray lines. DANTE DPP was compared to actual beamline DPP, XIA-XMAP, in two different XRF and XAFS experiments.

\subsection{Experimental setup and procedure}
\label{sec:LUCIAsetup}
\begin{figure}[htb!]
\centering
\includegraphics[width=90mm]{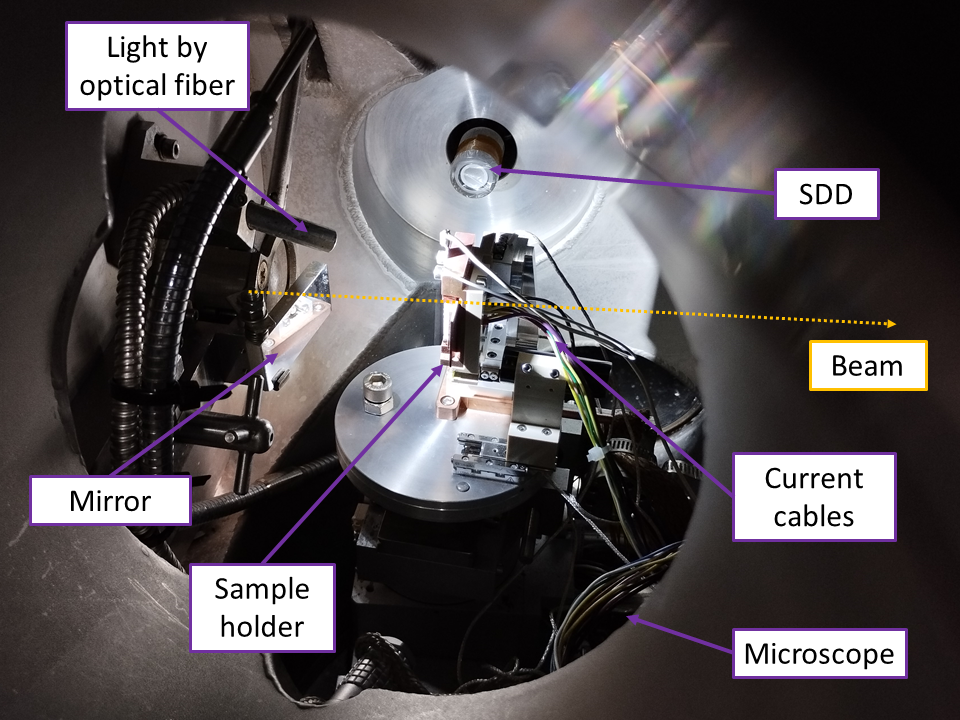}
\caption{Photo of the experimental setup at LUCIA beamline.}
\label{fig:LUCIASchemaTest}
\end{figure}

In LUCIA beamline, fluorescence data is collected by a SDD detector monoelement (BRUKER, model XFlash 6|60), installed at 90~degrees from the incident beam, as shown in Fig.~\ref{fig:LUCIASchemaTest}. The SDD is equipped with a polymer thin window so that fluorescence of elements down to the Carbon can be observed. Detector specifications are shown in Table~\ref{tab:SDDFeatures}. The detector ICR is adjusted by tilting the sample surface towards the detector for angles between 0 to 45~degrees and by varying the detector-to-sample distance between few millimeters and 25~cm. The beam intensity is monitored by measuring with a picoammeter the total drain current of a polymer film coated with 10~nm-thick of Nickel, located at the exit of the KB mirror vessel.

\begin{itemize}
    \item In XRF experiment, we tested the capability of DANTE in discriminating closely spaced fluorescence lines, as previously done in~\cite{Bombelli2019TowardsOX} but in a large ICR range. A homemade glass sample (Na$_2$O-MgO-2SiO$_2$) was analyzed in fluorescence mode at an excitation energy of 2.4~keV. The detector ICR was varied between 10 and 1800~kcps and energy spectra of 30~s exposure time were acquired. DANTE DPP was alternatively operated with LE firmware, at PT between 64 and 2048~ns; and with HR firmware, at a minimum PT of 64~ns and a maximum PT of 20484~ns. The same offline analysis described in Sec.~\ref{sec:labexp} was used to estimate the energy resolution (FWHM) of all lines.
    \item In XAS experiment, an iron (Fe) oxyhydroxide sample was analyzed to test DANTE at the energy range of the Fe K-edge (7212~eV). The spectra were collected in fluorescence mode between 7000 and 7200~eV with a step of 1~eV in the pre-edge region (7100 and 7110~eV), 0.2~eV in the X-ray absorption near-edge spectroscopy (XANES) region (7110-7160~eV) and 1~eV in the first part of the extended X-ray absorption fine structure (EXAFS) region (7160-7200~eV). For each point, a counting time of 1~s was used. Spectra were acquired at five detector ICR values (100, 300, 500, 700 and 1000~kcps), and alternating DANTE and XIA-XMAP DPPs. DANTE was operated with LE firmware at a PT of 496~ns. In the offline analysis, the XAS spectra were first corrected by the detector dead time and then normalized by Athena software~\cite{Ravel:ph5155} in order to compare their shape
\end{itemize}

\subsection{Results and discussion}
\label{sec:LUCIAresults}
\begin{figure}[htb!]
\centering
\includegraphics[width=75mm]{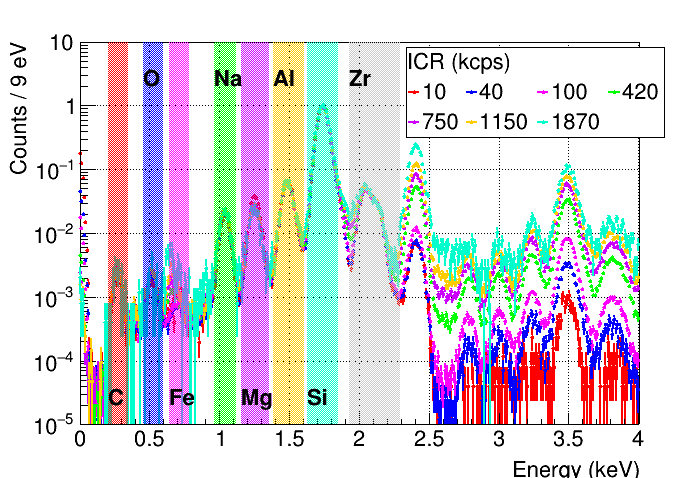}
\includegraphics[width=75mm]{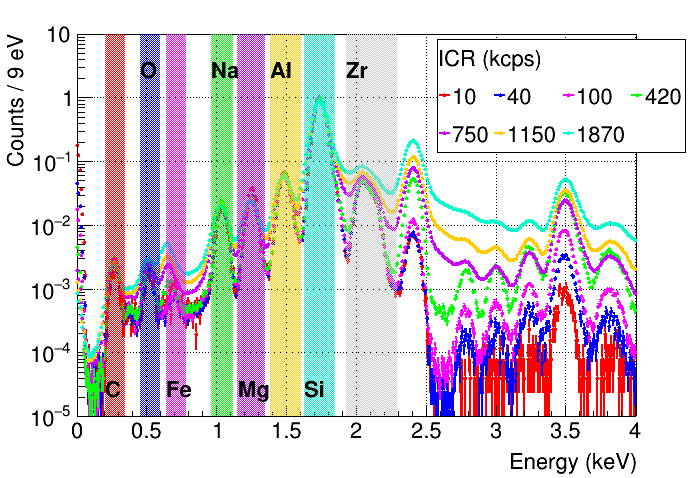}
\caption{Energy specrum of a glass target at ICR values between 10 and 1800~kcps measured by BRUKER SDD and acquired by DANTE DPP operated with LE firmware at a PT of 1024~ns (left) and HR firmware (right) at LUCIA beamline. Energy spectra have been normalized to the maximum for a better study of their shape. The different fluorescence lines are marked by shaded areas in the spectrum.}
\label{fig:LUCIAXRFGlass}
\end{figure}

XRF energy spectra of the glass sample are shown in Fig.~\ref{fig:LUCIAXRFGlass}, for LE and HR firmwares.
They are composed of the beam elastic peak (at 2.4~keV) and different sample fluorescence X-ray lines: carbon K-line (at 0.277~keV), oxygen K-line (at 0.525~keV), sodium K-line (at 1.041~keV), magnesium K-line (at 1.254~keV), aluminium K-line (at 1.497~keV), silicon K-line (at 1.74~keV) and zirconium L-lines (at 2.04~keV and 2.12~keV). Another line, situated at $\sim$0.7~keV, may be generated by the iron present in the sample environment. Peaks situated at energies above 2.4~keV are pile-up lines.

For all spectra, DANTE can discriminate lines event at ICR values of 1800~kcps. In the spectra of LE firmware, the peak shape remains the same, which indicates that energy resolution is constant. However, at high ICR values, energy spectra start lacking of statistics, which points to a high dead time. In the spectra of HR firmware, the shape deforms slightly at high ICR, which indicates a degradation of energy resolution, but the throughput is higher, i.e, dead time remains moderate.

\begin{figure}[htb!]
\centering
\includegraphics[width=75mm]{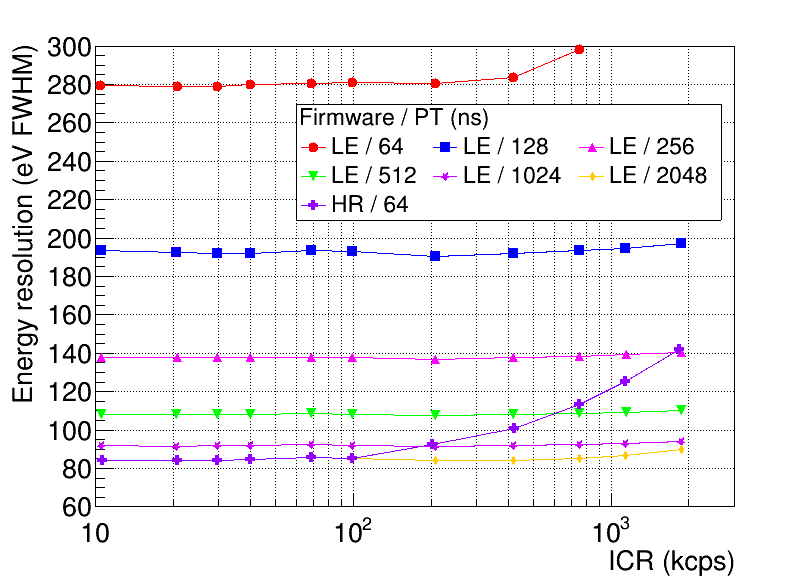}
\includegraphics[width=75mm]{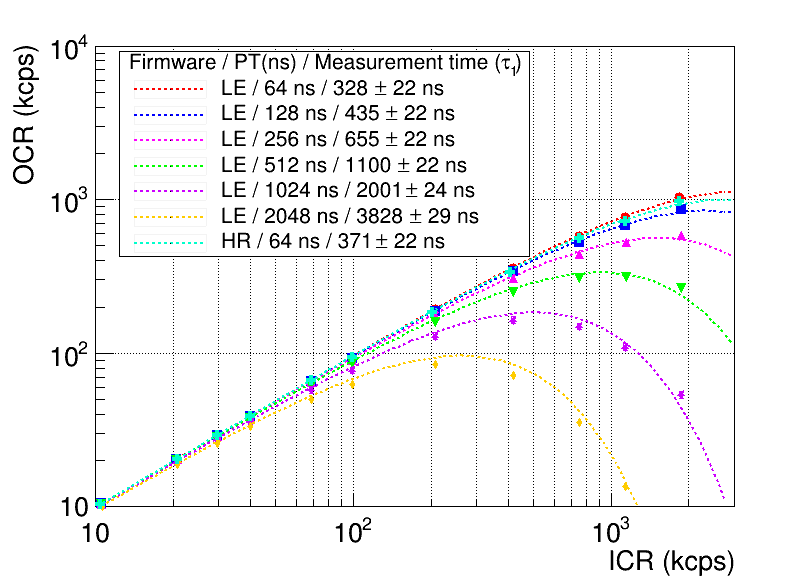}
\caption{Dependence of the energy resolution (FHWM) at 1.74~keV vs ICR (left); and the linearity of OCR vs ICR (right), for DANTE DPP with LE and HR firmwares and different PT settings.}
\label{fig:LUCIAXRFRes}
\end{figure}

These two dependencies (energy resolution at 1.74~keV and OCR) with ICR are quantified in Fig.~\ref{fig:LUCIAXRFRes}. As shown before, the energy resolution remains constant for ICR values between 10 and 1800~kcps for LE firmware and the specific value depends on the PT, as previously shown in the laboratory tests in Sec.~\ref{sec:labreslef}. The best value for energy resolution (84.2$\pm$0.2~eV) has been measured at a PT of 2048~ns. For HR firmware, the energy resolution degrades with ICR, from the optimum value at ICR of 10-100~kcps up to 141.8$\pm$0.2~eV at 1800 kcps. Meanwhile, HR firmware provides a high OCR (measurement time of 371~ns), not far from that of LE firmware at a PT of 64~ns (328~ns).

\begin{figure}[htb!]
\centering
\includegraphics[width=75mm]{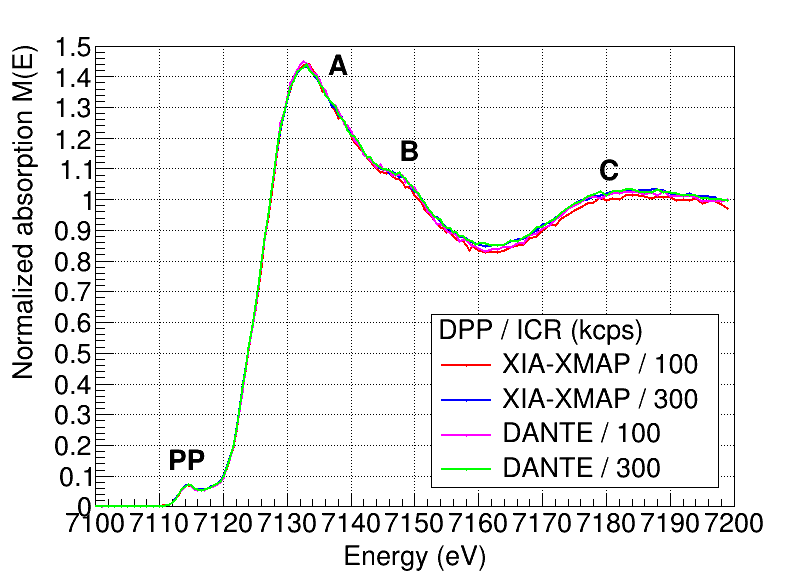}
\includegraphics[width=75mm]{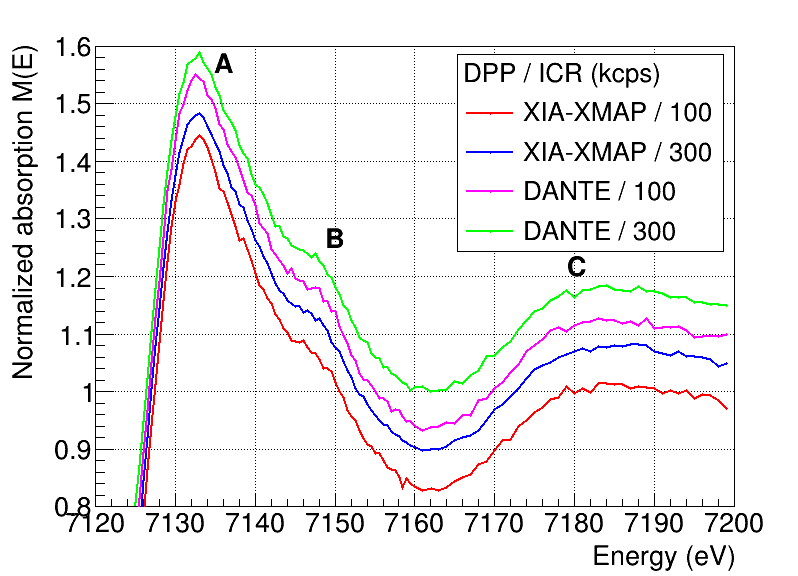}
\caption{Energy spectra (global view on the left, zoom on the right) at Fe K-edge of a Fe oxyhydroxide sample measured by BRUKER SDD and acquired by XIA-XMAP and DANTE DPPs at detector ICR of 100, 300, 500, 700 and 1000~kcps at LUCIA beamline. Global view spectra of DANTE DPP have been shifted 0.1 units for a better comparison to XIA-XMAP spectra. Zoomed spectra have a successive vertical offset of 0.05 units for a better comparison of their shape.}
\label{fig:LUCIAXAFSFerri}
\end{figure}

The XAS spectra of the Fe oxyhydroxide at the Fe K-edge are shown in Fig.~\ref{fig:LUCIAXAFSFerri}. The spectrum is dominated by a main peak at 7133~eV followed by a shoulder at $\sim$7150~eV (labeled A and B). These two features are characteristic of the electronic structure of the iron atom in the Fe oxyhydroxide structure and give information on the oxidation state (position of the white line) and the structural environment (shape of the spectrum) of the iron. The broad peak after the XANES part around 7180~eV (labeled C) is the first oscillation of the EXAFS part. EXAFS provides information on the interatomic distances between the excited atom (Fe in our case) and his neighbors. We can also notice in the pre-edge region a pre-peak (labeled PP) around 7115~eV (Fig.~\ref{fig:LUCIAXAFSFerri}, left). This weak peak is a powerful tool to determine the oxidation state of the excited atom, as shown in many papers~\cite{Wilke2001mw,FARGES2004176,WILKE200471}.

The energy spectra generated by XIA-XMAP and DANTE DPPs have similar shape and show the expected structure for a Fe oxyhydroxide sample. We can conclude that both XIA-XMAP and DANTE can be operated in XAS experiments at ICR values up to 1000~kcps. The only difference appears at high ICR values (700-1000~kcps), where a little deformation of the main peak is observed for XIA-XMAP spectra (0.074 in absolute units), while the deformation is insignificant for DANTE spectra (0.015 in absolute units).

\section{XRF cartography at PUMA beamline}
\label{sec:PUMAint}
PUMA beamline of SOLEIL synchrotron is a hard X-ray imaging beamline dedicated to the study of materials from cultural heritage which uses, among others, the XRF technique~\cite{Tack:ve5108}. The DPP used to acquired XRF data requires to perform on-the-fly measurements while scans are made, at exposure times between 20-200~ms and keeping an optimum energy resolution and moderate dead time values. This ability was demonstrated for DANTE and compared to actual beamline DPP, XIA-FalconX.

\subsection{Experimental setup and procedure}
\label{sec:PUMAsetup}
\begin{figure}[htb!]
\centering
\includegraphics[width=90mm]{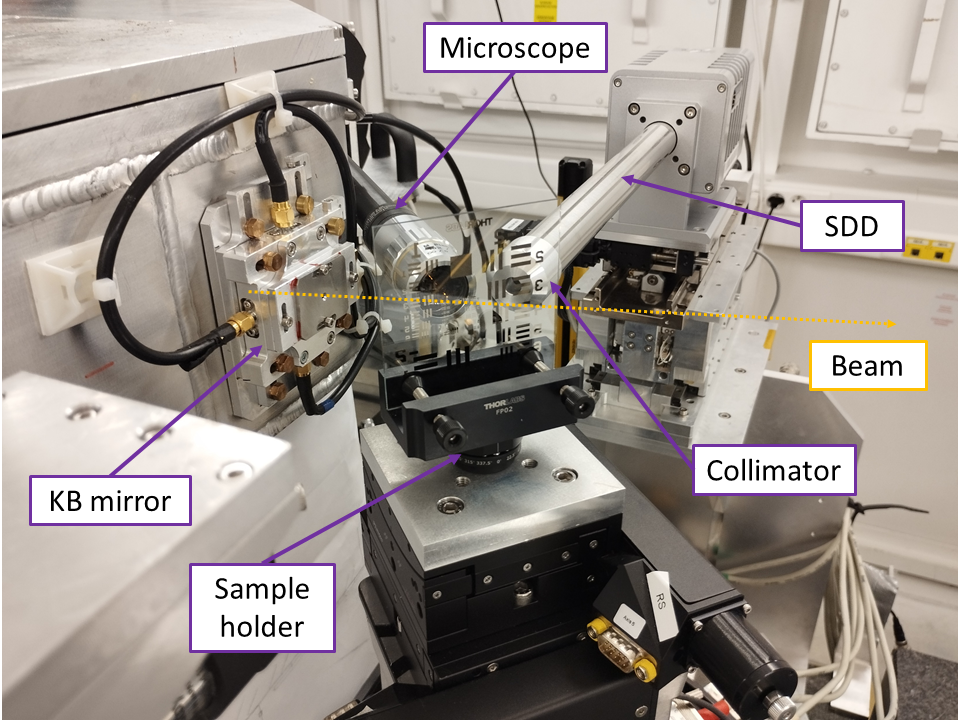}
\caption{Photo of the experimental setup at PUMA beamline,
with an alignment target on the sample holder. The KB mirror focuses the X-rays on the sample, whose surface is oriented at 45$^\circ$ to the beam. The microscope visualize the sample and allows the alignment with respect to the beam. The SDD, equipped with a collimator, is mounted at 90$^\circ$ to the incident beam.}
\label{fig:PUMASchemaTest}
\end{figure}

XRF measurements were performed using an excitation energy of 18~keV.
The beam was focused on the sample by a KB mirror
(see Fig.~\ref{fig:PUMASchemaTest}), leading to an effective beam footprint of 5~$\mu$m (FWHM) in vertical and 7~$\mu$m (FWHM) in horizontal direction.
XRF data were acquired by a single-element SDD (RAYSPEC, model SiriusSD-100140C-Be), installed at 90~degrees from the incident beam. Detector specifications are shown in Table~\ref{tab:SDDFeatures}. A cylindrical-shaped aluminium collimator with 8 mm internal diameter and 10~mm length was used to reduce the spectral contribution of background scattering and unwanted signals from the sample environment. A visible light microscope (Hamamatsu digital camera C11440), equipped with a $10\times$ magnification lens and located perpendicularly to the sample surface was used to monitor the sample position and to assure its position in the beam focal point.

During XRF analysis, a pigment reference sample was moved continuously through the beam in X and Z-axis over an area of 1~mm $\times$ 2~mm at a given speed and 100 $\times$ 200 spectra were acquired with flyscan system~\cite{Medjoubi:vv5054} in intervals of 10~$\mu$m and an exposure time of 100~ms. Scans were made at two different detector-to-sample distances (70 and 90~mm), and alternating DANTE and XIA-FalconX DPPs. DANTE DPP was operated with high-rate (HR) firmware, at a minimum PT of 32~ns and a maximum PT of 2016~ns. XIA-FalconX DPP was operated with the release 20.4 of firmware and a calibration file optimized for the SDD detector (very short decay time, low rate filter and pulse characterization at an analog gain of 2.0). XRF spectra were calibrated and integrated using PyMCA software~\cite{SOLE200763} before normalization for incident beam flux variation and DPP dead time.

\subsection{Results and discussion}
\label{sec:PUMAresults}
\begin{figure}[htb!]
\centering
\begin{tabular}{ccccc}
    & \multicolumn{2}{c}{Input Count Rate (kcps)} & \multicolumn{2}{c}{Dead time (\%)}\\
     & 70 mm & 90 mm & 70 mm & 90 mm\\
    \rotatebox[origin=c]{90}{DANTE} &
    \adjustimage{width=.21\textwidth,valign=c}{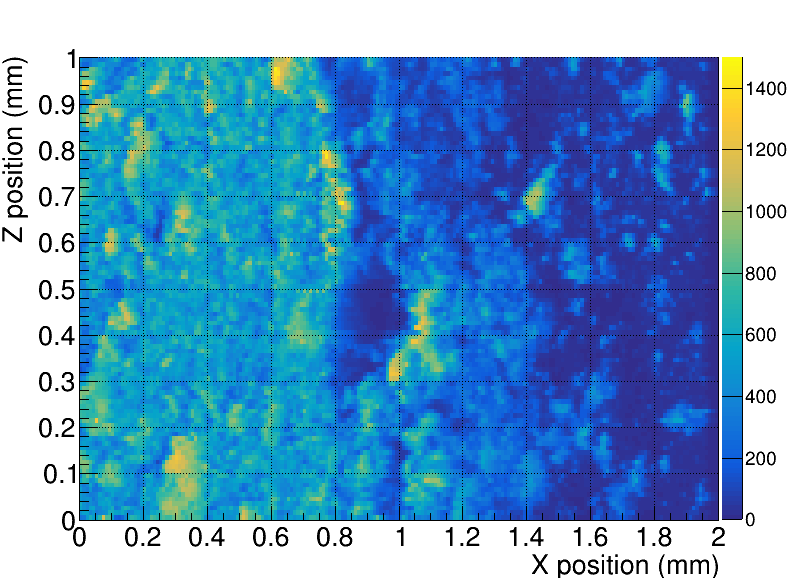}&
    \adjustimage{width=.21\textwidth,valign=c}{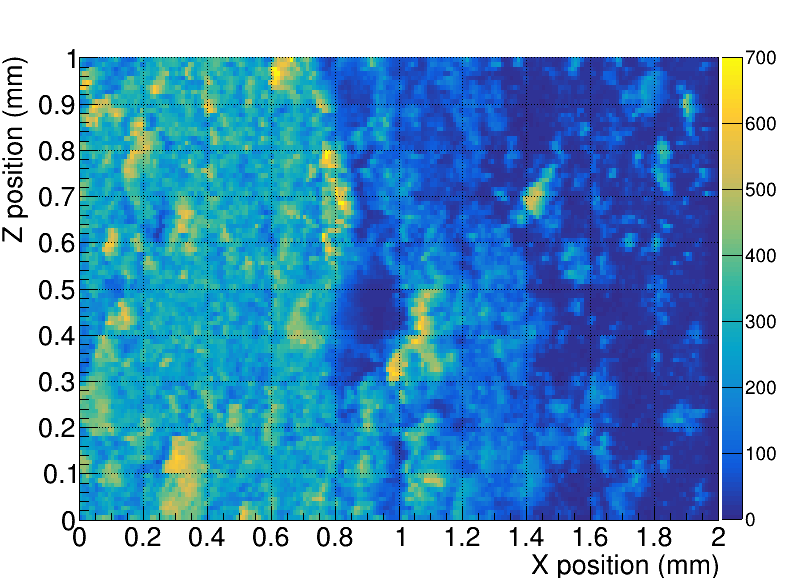}&
    \adjustimage{width=.21\textwidth,valign=c}{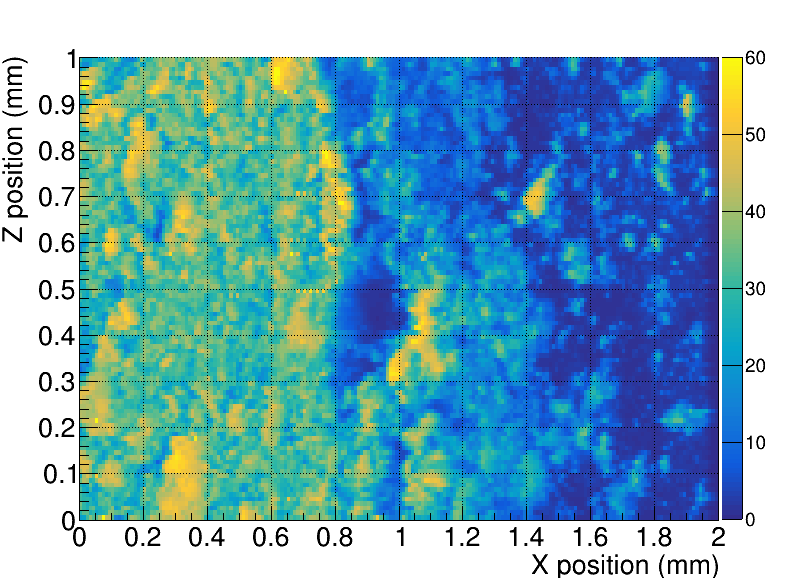}&
    \adjustimage{width=.21\textwidth,valign=c}{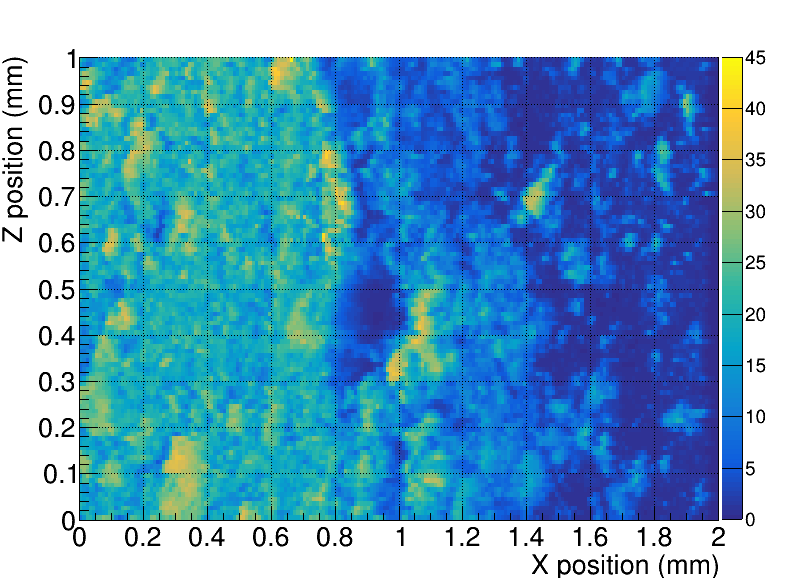}\\
    \rotatebox[origin=c]{90}{FalconX}&
    \adjustimage{width=.21\textwidth,valign=c}{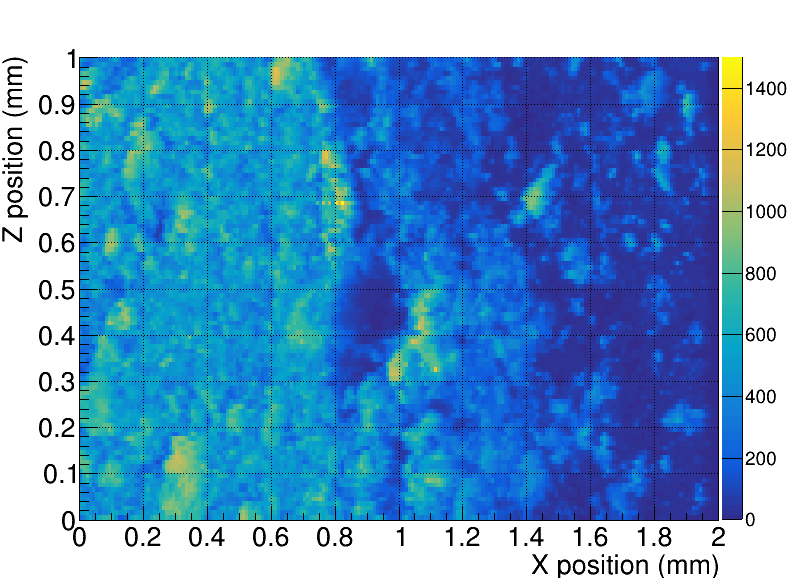}&
    \adjustimage{width=.21\textwidth,valign=c}{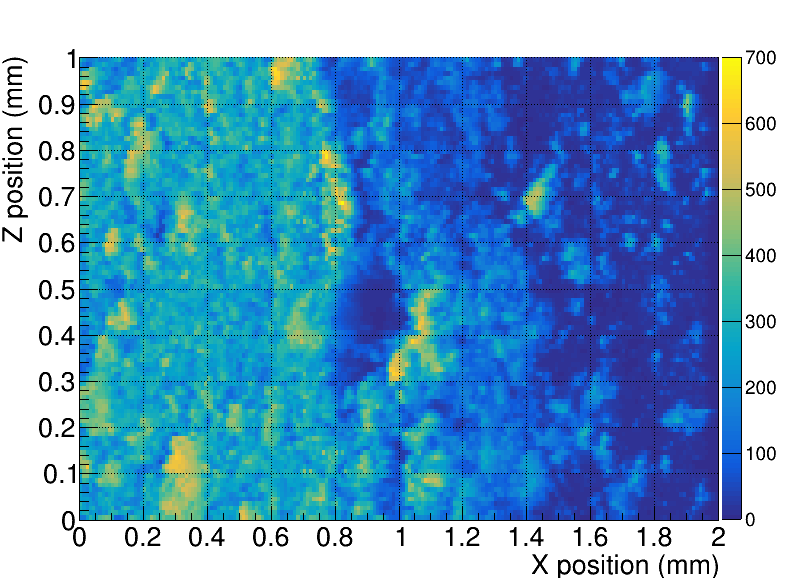}&
    \adjustimage{width=.21\textwidth,valign=c}{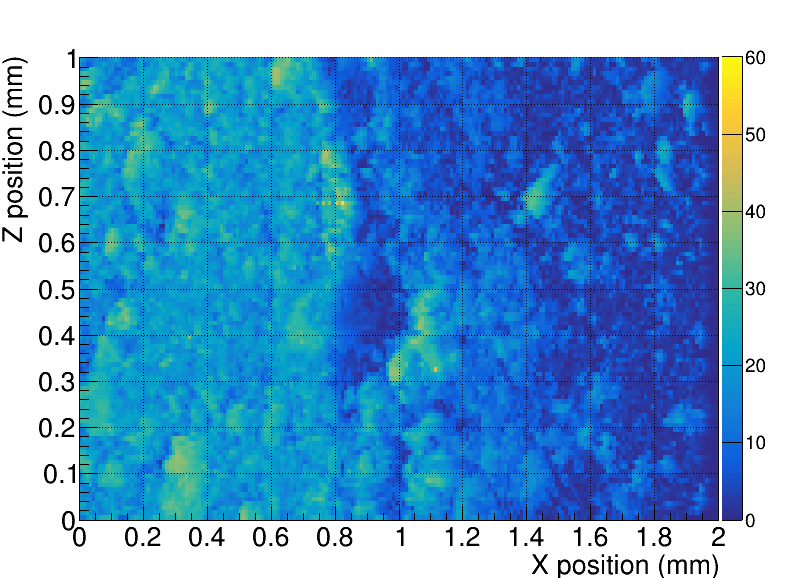}&
    \adjustimage{width=.21\textwidth,valign=c}{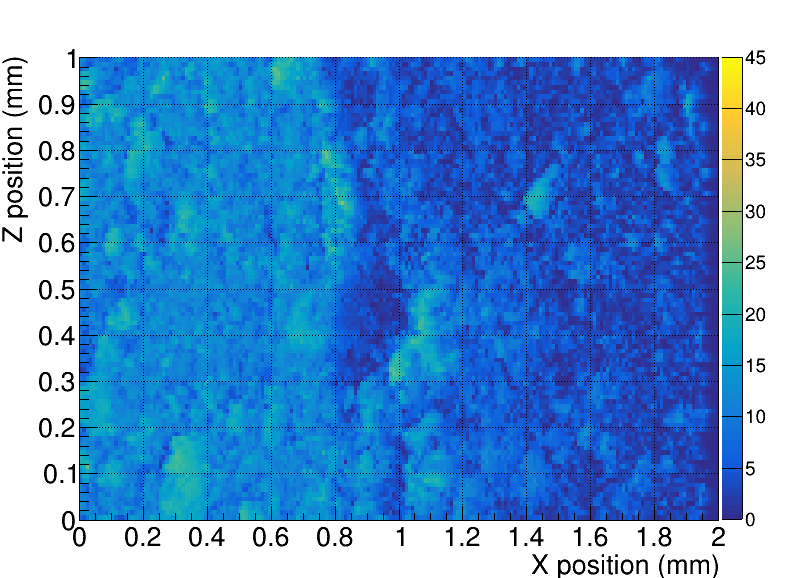}\\
 \end{tabular}
\caption{Input Count Rate and dead time maps for a pigment reference stone sample over an area of 1~mm $\times$ 2~mm, for DANTE and XIA-FalconX DPPs and for detector-to-sample distances of 70~mm and 90~mm.}
\label{fig:PUMAICRDT}
\end{figure}

The distributions of the ICR and dead time (DT) are respectively shown in Fig.~\ref{fig:PUMAICRDT}. DANTE DPP shows around 50\% higher DT values than XIA-FalconX for spectra with maximum ICR (740 and 1450~kcps, respectively for the detector-to-sample distances of 70 and 90~mm), which is in agreement with previous results in Sec.~\ref{sec:compdpp} for HR firmware and VORTEX-4EM SDD detector (DT of 28.8\% vs 17.6\% at 1~Mcps, Table \ref{tab:DANTEComp}). This difference has been later confirmed for RAYSPEC SDD and the X-ray generator source.

\begin{figure}[htb!]
\centering
\begin{tabular}{cccc}
     & Ca & Mn & Fe\\
    \rotatebox[origin=c]{90}{DANTE} &
    \adjustimage{width=.29\textwidth,valign=c}{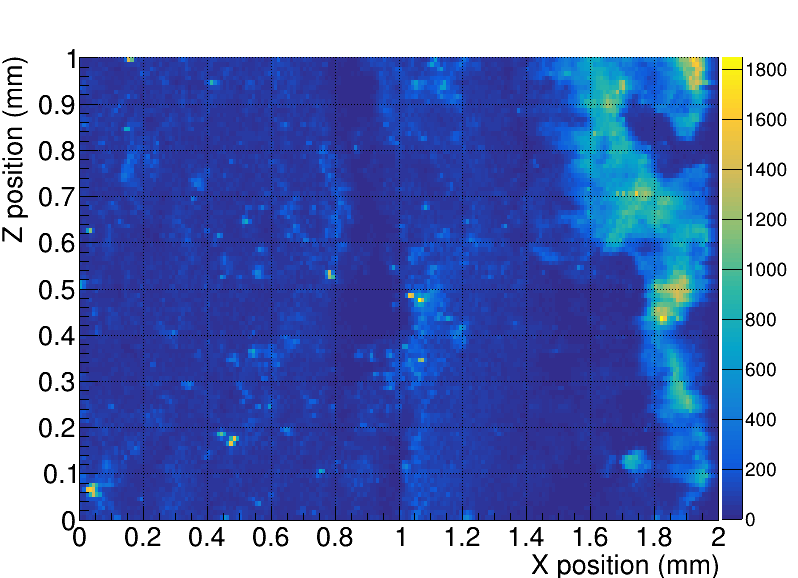}&
    \adjustimage{width=.29\textwidth,valign=c}{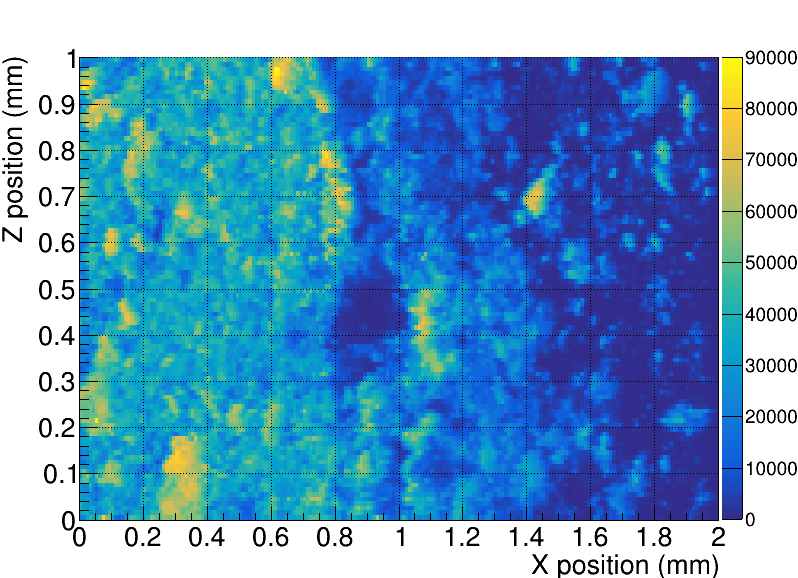}&
    \adjustimage{width=.29\textwidth,valign=c}{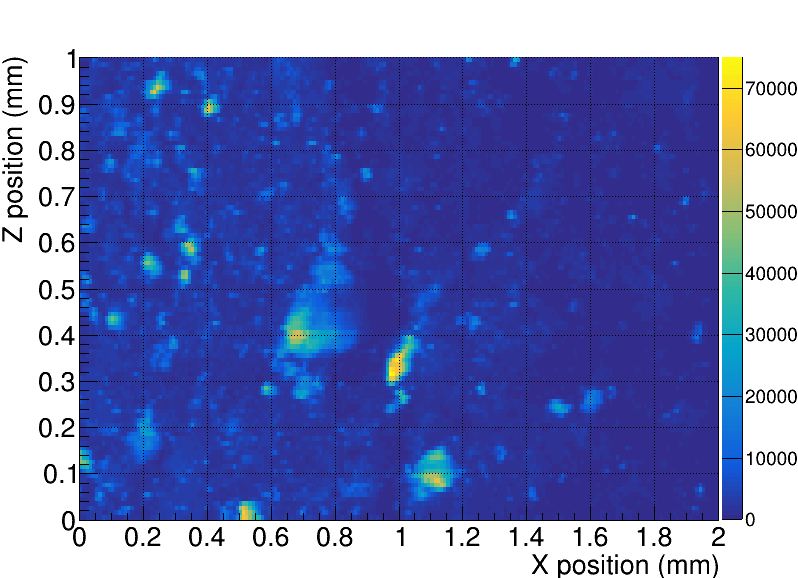}\\
    \rotatebox[origin=c]{90}{XIA-FalconX}&
    \adjustimage{width=.29\textwidth,valign=c}{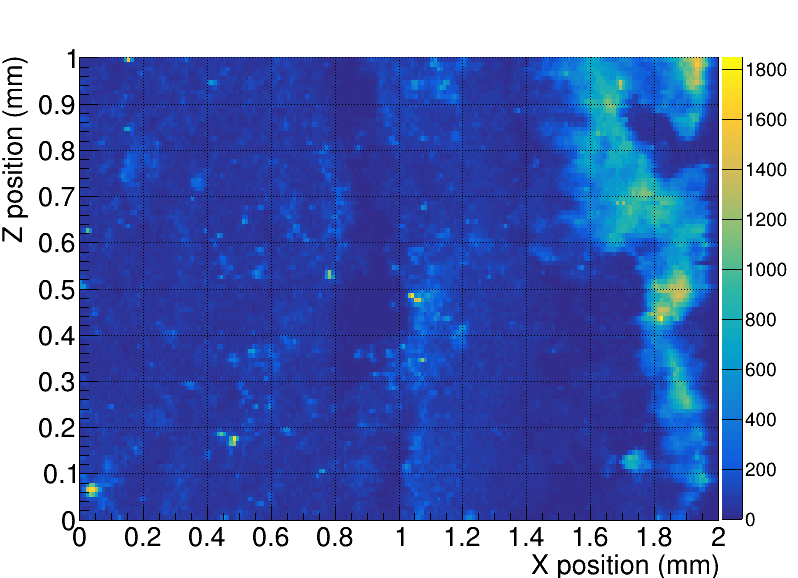}&
    \adjustimage{width=.29\textwidth,valign=c}{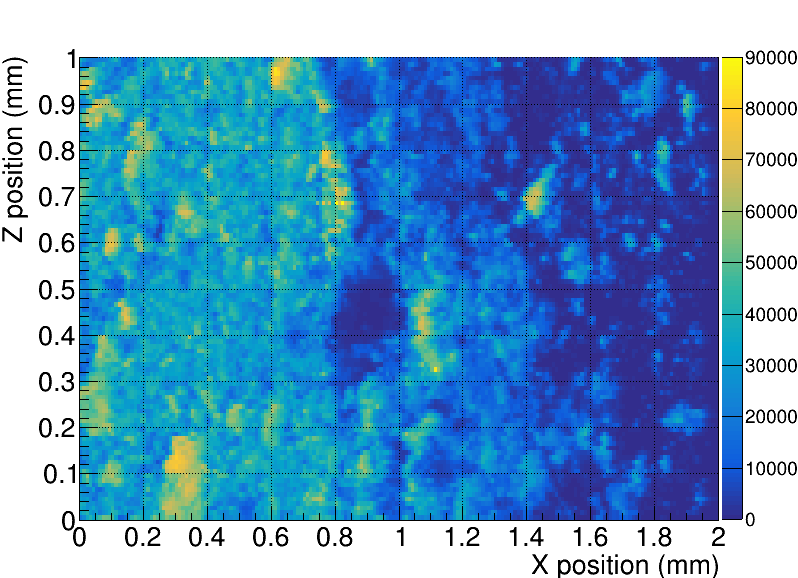}&
    \adjustimage{width=.29\textwidth,valign=c}{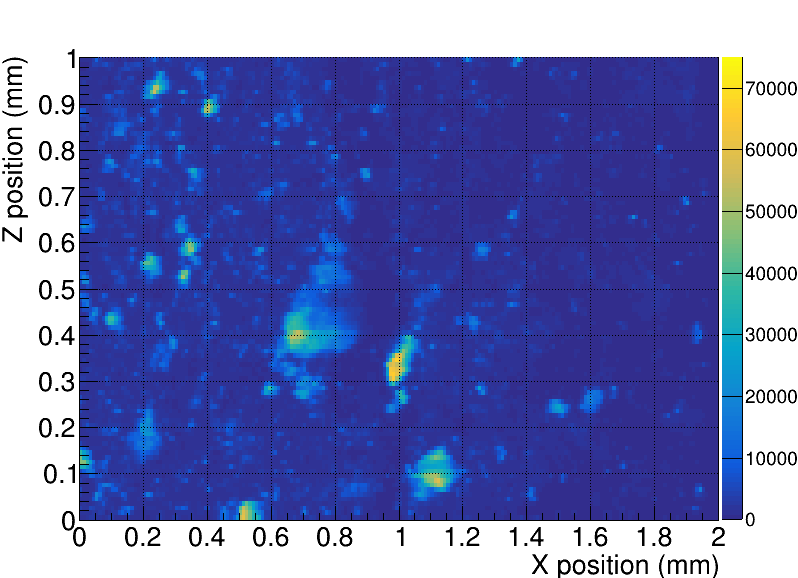}\\
     & As & Br & Ba\\
    \rotatebox[origin=c]{90}{DANTE} &
    \adjustimage{width=.29\textwidth,valign=c}{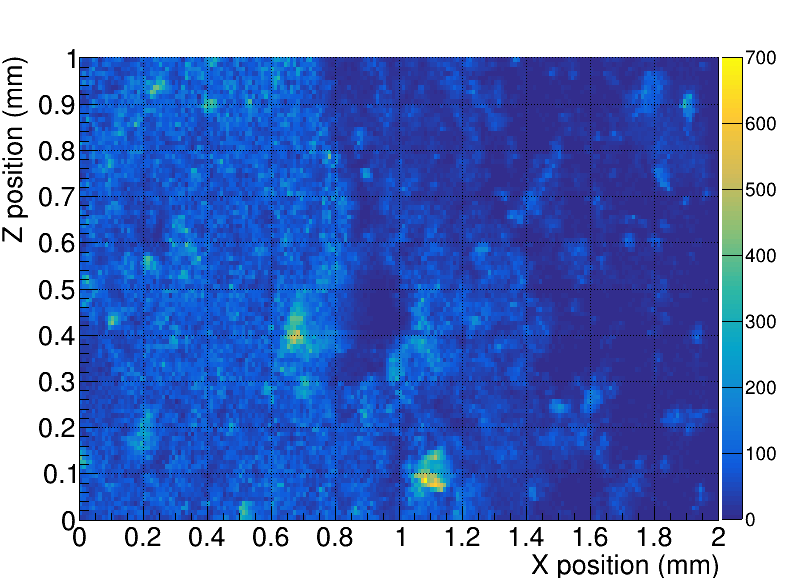}&
    \adjustimage{width=.29\textwidth,valign=c}{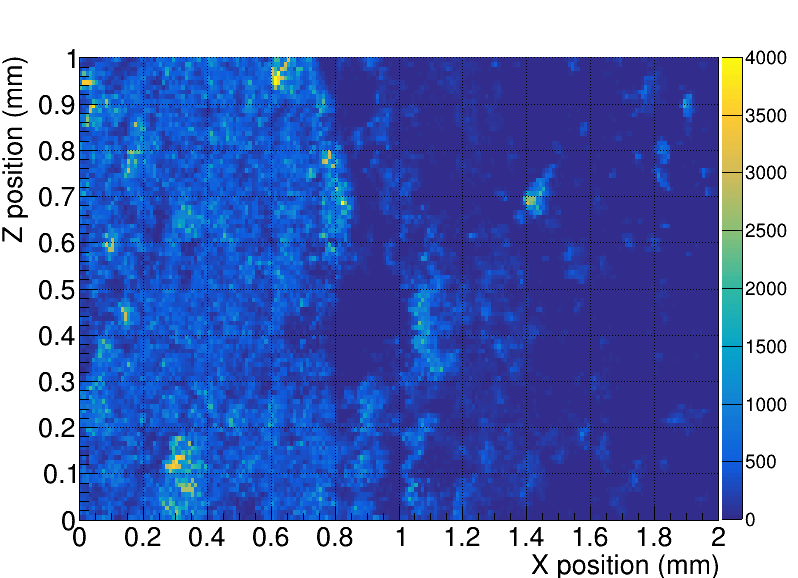}&
    \adjustimage{width=.29\textwidth,valign=c}{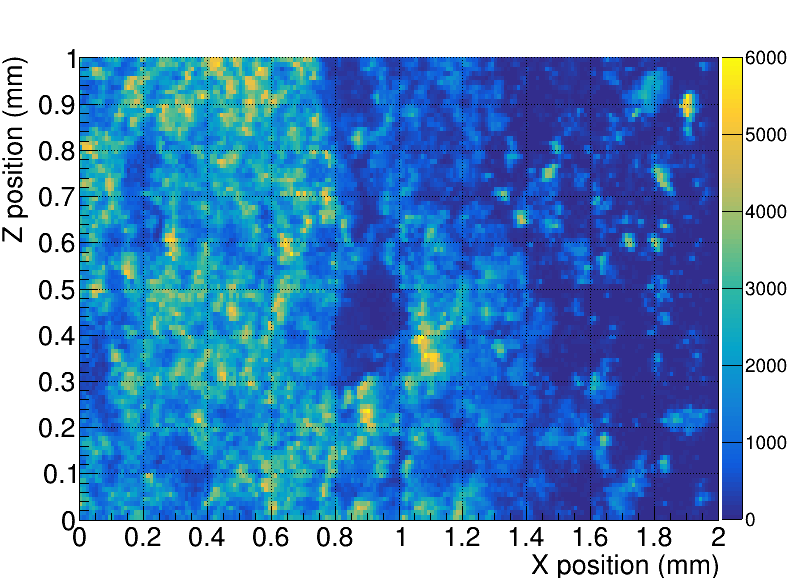}\\
    \rotatebox[origin=c]{90}{XIA-FalconX}&
    \adjustimage{width=.29\textwidth,valign=c}{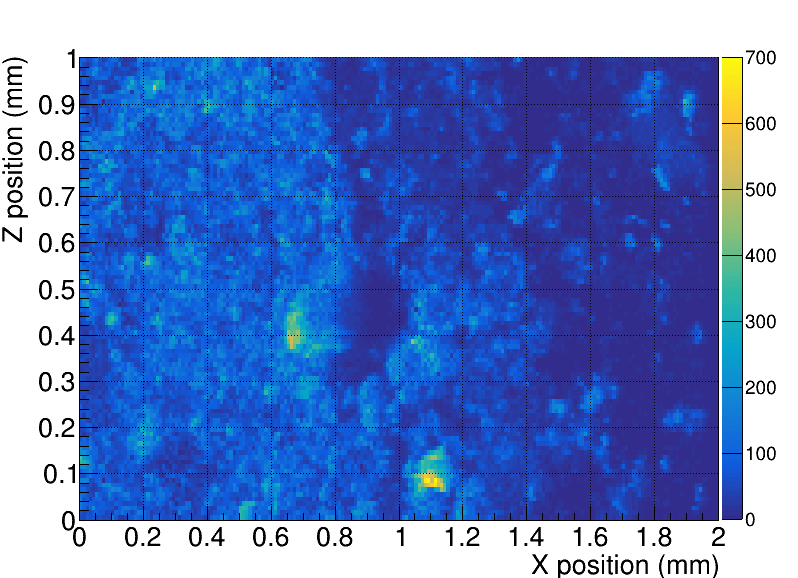}&
    \adjustimage{width=.29\textwidth,valign=c}{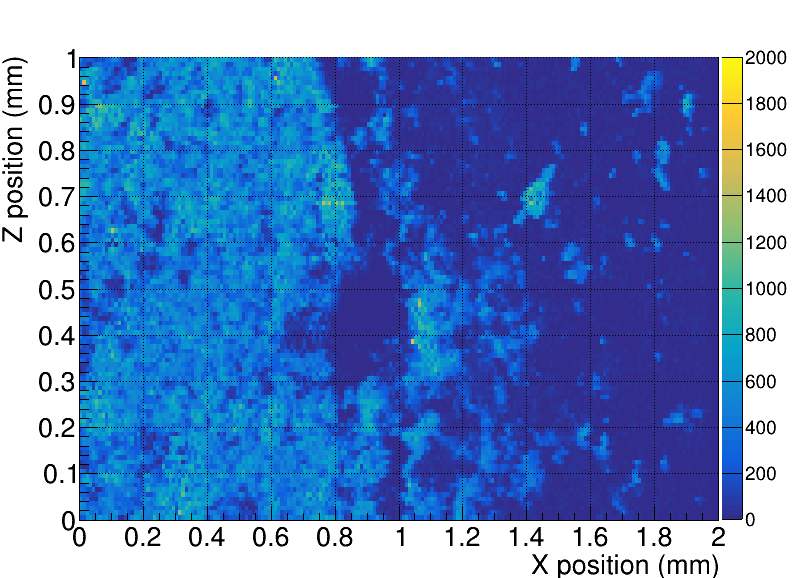}&
    \adjustimage{width=.29\textwidth,valign=c}{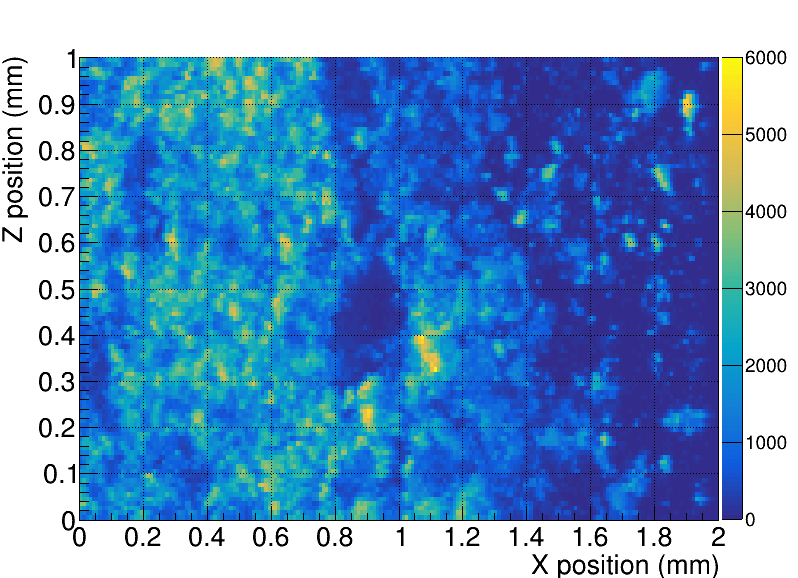}\\
\end{tabular}
\caption{XRF maps of Ca, Mn and Fe (top) and of As, Br and Br (bottom) of a pigment reference sample over an area of 1~mm $\times$ 2~mm, for DANTE and XIA-FalconX DPPs and for a detector-to-sample distance of 70~mm, in XRF experiment at PUMA beamline.}
\label{fig:PUMAXRFmaps}
\end{figure}

\begin{figure}[htb!]
\centering
\includegraphics[width=75mm]{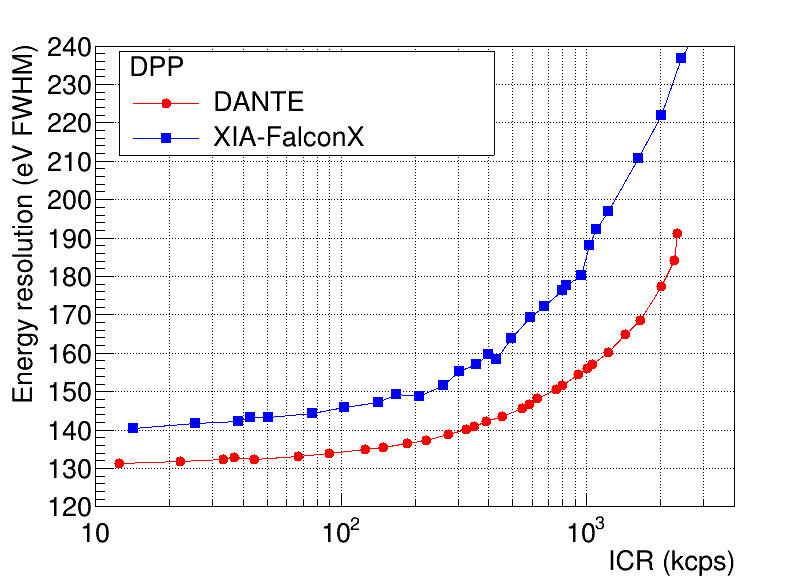}
\includegraphics[width=75mm]{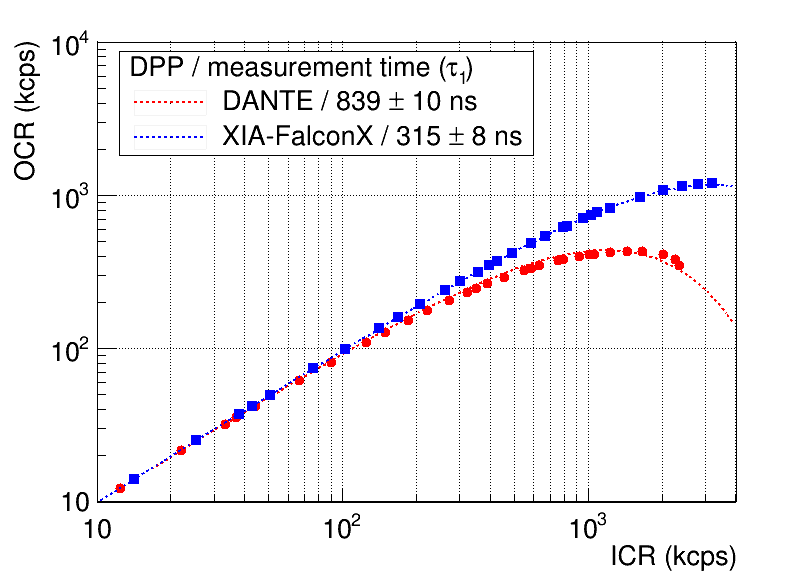}
\caption{Dependence of the energy resolution (FHWM) at 5.9~keV vs ICR (left); and the linearity of OCR vs ICR (right), measured by RAYSPEC SDD and read by DANTE and XIA-FalconX DPPs. Dashed lines of the right plot are the fit result of expression~\ref{eq:tau1} to experimental data.}
\label{fig:PUMAXRFRes}
\end{figure}

In Fig.~\ref{fig:PUMAXRFmaps}, the distribution of the main components of the pigment reference sample are shown for the two DPPs at a detector-to-sample distance of 70~mm. Comparing the chemical maps, no clear difference in terms of ICR contrast or spatial resolution is observed between the images provided by the two DPPs. Only slightly higher maximum output throughput is provided by DANTE DPP, ranging from 5\% for Calcium to 15\% for Iron. This fact can be explained by the better energy resolution of DANTE DPP at high ICR compared to FalconX (as previously observed in laboratory tests with the X-ray generator source, see Fig.~\ref{fig:PUMAXRFRes}, left), which reduces the loss of peak integral due to peak tails. This result gives evidence that DANTE DPP with HR firmware can be used in XRF experiments in hard X-ray regime, showing a similar performance than those of XIA-FalconX.

\section{Conclusions}
\label{sec:conc}
We have evaluated the performance of DANTE Digital Pulse Processor, with the selection of Low-Energy optimized (LE) and High-Rate optimized (HR) firmwares and a wide range of peaking time values, with an X-ray generator at the laboratory and two different SDD detectors. Its performance for Input Count Rate (ICR) between 10~kcps and 1~Mcps has been characterized in terms of energy resolution, peak stability, measurement time (i.e. dead time) and time resolution (i.e. pile-up inspection).

For LE firmware, energy resolution measurements by DANTE are compatible with those made by XIA-XMAP DPP at 10~kcps and for PT values longer than 1000~ns, while results are slightly worse at shorter PT values. From an ICR of 300~kcps, DANTE reaches an optimum energy resolution at PT values longer than 1000~ns, where XIA-XMAP DPP becomes saturated. Peak shifts for DANTE are smaller than 0.1-0.3\% and the estimated values for time resolution (54-67~ns) are much better than those measured by XIA-XMAP (166-271~ns).

For HR firmware, the energy resolution is optimum at ICR values between 10 and 100~kcps, and it progressively degrades at higher values. Nevertheless, the energy resolution is better than that of LE firmware, while DANTE provides a higher Output Count Rate (and smaller dead time values). Peak shifts are smaller than 0.05-0.15\% and the estimated values for time resolution (50-68~ns) remain optimal. Compared to other recent DPP in the market (XIA-FalconX and Xspress3), DANTE shows a longer measurement time (i.e., a higher dead time) but a shorter time resolution (i.e., a reduced pile-up ratio).

DANTE has also been tested in three different experiments at SOLEIL synchrotron. In an XRF experiment at LUCIA beamline, we have demonstrated the ability of DANTE in discriminating closely spaced fluorescence soft X-ray lines (200-2000~eV) of a glass sample for ICR values between 10 and 1800~kcps using a homemade glass sample. DANTE has also shown a good performance in an XAS experiment with a Fe oxyhydroxide sample at Fe K-edge ($\sim$7~keV), comparable to that of XIA-XMAP DPP, and for ICR values between 100 and 1000~kcps. Finally, DANTE performance has been evaluated in a hard X-ray XRF cartography experiment at PUMA beamline using a pigment reference sample. Chemical maps generated by DANTE show the same contrast and spatial resolution as those provided by XIA-FalconX.

DANTE is an operational DPP at SOLEIL synchrotron. In future XAS experiments, DANTE will be operated at higher ICR values to improve the S/N ratio; and in XRF experiments, at exposures time of 1-10~ms to improve the spatial resolution of chemical maps.

\acknowledgments
The authors acknowledge SOLEIL for provision of synchrotron radiation facilities (proposal number 20191274) at LUCIA and PUMA beamlines. The authors would like to thank SOLEIL computing control service (ISAC), in particular to C.~Castel (Mediane Système), A. Hercule (Thales Services Numériques), F.~Langlois and A.~Noureddine, for their work in the integration of DANTE API library in TANGO SOLEIL data-flow system.

\bibliographystyle{JHEP}
\bibliography{20230511_FJIguaz_DANTEArticle}

\providecommand{\href}[2]{#2}\begingroup\raggedright\begin{thebibliography}{10}

\bibitem{grieken2001xrf}
R.~Van~Grieken and A.~Markowicz, \emph{{Handbook of X-Ray Spectrometry}}.
\newblock CRC Press, 2001,
  \href{http://dx.doi.org/10.1201/9780203908709}{10.1201/9780203908709}.

\bibitem{calvin2013xafs}
S.~Calvin, \emph{{XAFS for Everyone}}.
\newblock CRC Press, 2013,
  \href{http://dx.doi.org/10.1201/b14843}{10.1201/b14843}.

\bibitem{SCHLOSSER2010270}
D.~Schlosser, P.~Lechner, G.~Lutz, A.~Niculae, H.~Soltau, L.~Str\"ader et~al.,
  \emph{{Expanding the detection efficiency of silicon drift detectors}},
  \href{http://dx.doi.org/https://doi.org/10.1016/j.nima.2010.04.038}{\emph{Nucl.
  Instrum. Meth. A} {\bfseries 624} (2010) 270--276}.

\bibitem{SANGSINGKEOW2003183}
P.~Sangsingkeow et~al., \emph{{Advances in germanium detector technology}},
  \href{http://dx.doi.org/https://doi.org/10.1016/S0168-9002(03)01047-7}{\emph{Nucl.
  Instrum. Meth. A} {\bfseries 505} (2003) 183 -- 186}.

\bibitem{Bordessoule2019}
M.~Bordessoule, E.~Fonda, N.~Guignot, J.~P. Itie, C.~Menneglier and F.~Orsini,
  \emph{{Performance of spectroscopy detectors and associated electronics
  measured at SOLEIL synchrotron}},
  \href{http://dx.doi.org/10.1063/1.5084701}{\emph{AIP Conference Proceedings}
  {\bfseries 2054} (2019) 060070}.

\bibitem{Heald:rv5031}
S.~M. Heald, \emph{{Strategies and limitations for fluorescence detection~of
  XAFS at high flux beamlines}},
  \href{http://dx.doi.org/10.1107/S1600577515001320}{\emph{Journal of
  Synchrotron Radiation} {\bfseries 22} (Mar, 2015) 436--445}.

\bibitem{XIAFalcon}
XIA, ``{FalconX8, Ultra-high rate Digital Pulse Processor using SITORO patented
  signal processing technology}.'' \url{https://xia.com/products/falconx/},
  Accessed: 2022-12-15.

\bibitem{Scoullar2011}
P.~A.~B. Scoullar, C.~C. McLean and R.~J. Evans, \emph{{Real Time Pulse
  Pile‐up Recovery in a High Throughput Digital Pulse Processor}},
  \href{http://dx.doi.org/10.1063/1.3665324}{\emph{AIP Conference Proceedings}
  {\bfseries 1412} (2011) 270--277}.

\bibitem{QuantumXspress3}
QuantumDetectors, ``{Xspress 3X/M}.''
  \url{https://quantumdetectors.com/products/xspress3/}, Accessed: 2022-12-15.

\bibitem{FARROW1995567}
R.~Farrow, G.~Derbyshire, B.~Dobson, A.~Dent, D.~Bogg, J.~Headspith et~al.,
  \emph{{XSPRESS — X-ray signal processing electronics for solid state
  detectors}},
  \href{http://dx.doi.org/10.1016/0168-583X(94)00370-X}{\emph{Nucl. Instr.
  Meth. A} {\bfseries 97} (1995) 567--571}.

\bibitem{XGLabDANTE}
XGLab, ``{DANTE: The Digital Pulse Processor for X-ray Spectroscopy}.''
  \url{xglab.it/products/dante/}, Accessed: 2022-12-15.

\bibitem{Bombelli2019TowardsOX}
L.~Bombelli, M.~Manotti, M.~Altissimo, G.~Kourousias, R.~Alberti and
  A.~Gianoncelli, \emph{{Towards on-the-fly X-ray fluorescence mapping in the
  soft X-ray regime}}, \href{http://dx.doi.org/10.1002/xrs.2998}{\emph{X-Ray
  Spectrometry} {\bfseries 48} (2019) 325--329}.

\bibitem{Pouyet2021ep}
E.~Pouyet, N.~Barbi, H.~Chopp, O.~Healy, A.~Katsaggelos, S.~Moak et~al.,
  \emph{{Development of a highly mobile and versatile large MA-XRF scanner for
  in situ analyses of painted work of arts}},
  \href{http://dx.doi.org/https://doi.org/10.1002/xrs.3173}{\emph{X-Ray
  Spectrometry} {\bfseries 50} (2021) 263--271}.

\bibitem{TANGO1999}
J.~Chaize, A.~G\"otz, W.~Klotz, J.~Meyer, M.~Perez and E.~Taurez, \emph{{TANGO
  - an object oriented control system based on CORBA}}, {\emph{Proceedings of
  ICALEPCS conference} (1999) 473--475}.

\bibitem{XIAXMAP}
XIA, ``{xMAP, 4 channel PXI Digital Pulse Processor with mapping features}.''
  \url{https://xia.com/support/xmap/}, Accessed: 2022-12-15.

\bibitem{Hubbard1996}
B.~Hubbard, W.~Warburton and C.~Zhou, \emph{{Digital X-ray processing
  electronics for fluorescence EXAFS and spectroscopy}},
  \href{http://dx.doi.org/10.1063/1.1147335}{\emph{Rev. sci. Instrum.}
  {\bfseries 67} (1996) 3371}.

\bibitem{6154396}
L.~Bombelli, C.~Fiorini, T.~Frizzi, R.~Alberti and A.~Longoni, \emph{{"CUBE", A
  low-noise CMOS preamplifier as alternative to JFET front-end for high-count
  rate spectroscopy}},
  \href{http://dx.doi.org/10.1109/NSSMIC.2011.6154396}{\emph{2011 IEEE Nuclear
  Science Symposium Conference Record} (2011) 1972--1975}.

\bibitem{6551138}
L.~Bombelli, C.~Fiorini, T.~Frizzi, R.~Alberti and R.~Quaglia, \emph{{High rate
  X-ray spectroscopy with “CUBE” preamplifier coupled with silicon drift
  detector}}, \href{http://dx.doi.org/10.1109/NSSMIC.2012.6551138}{\emph{2012
  IEEE Nuclear Science Symposium and Medical Imaging Conference Record
  (NSS/MIC)} (2012) 418--420}.

\bibitem{Brun:1997pa}
R.~Brun and F.~Rademakers, \emph{{ROOT: An object oriented data analysis
  framework}},
  \href{http://dx.doi.org/10.1016/S0168-9002(97)00048-X}{\emph{Nucl. Instrum.
  Meth. A} {\bfseries 389} (1997) 81--86}.

\bibitem{FLANK2006269}
A.-M. Flank, G.~Cauchon, P.~Lagarde, S.~Bac, M.~Janousch, R.~Wetter et~al.,
  \emph{{LUCIA, a microfocus soft XAS beamline}},
  \href{http://dx.doi.org/https://doi.org/10.1016/j.nimb.2005.12.007}{\emph{Nuclear
  Instruments and Methods in Physics Research Section B: Beam Interactions with
  Materials and Atoms} {\bfseries 246} (2006) 269--274}.

\bibitem{Vantelon:vv5122}
D.~Vantelon, N.~Trcera, D.~Roy, T.~Moreno, D.~Mailly, S.~Guilet et~al.,
  \emph{{The LUCIA beamline at SOLEIL}},
  \href{http://dx.doi.org/10.1107/S1600577516000746}{\emph{Journal of
  Synchrotron Radiation} {\bfseries 23} (Mar, 2016) 635--640}.

\bibitem{Ravel:ph5155}
B.~Ravel and M.~Newville, \emph{{{\it ATHENA}, {\it ARTEMIS}, {\it HEPHAESTUS}:
  data analysis for X-ray absorption spectroscopy using {\it IFEFFIT}}},
  \href{http://dx.doi.org/10.1107/S0909049505012719}{\emph{Journal of
  Synchrotron Radiation} {\bfseries 12} (Jul, 2005) 537--541}.

\bibitem{Wilke2001mw}
M.~Wilke, F.~Farges, P.-E. Petit, G.~E. Brown and F.~Martin, \emph{{Oxidation
  state and coordination of Fe in minerals: An Fe K-XANES spectroscopic
  study}}, \href{http://dx.doi.org/doi:10.2138/am-2001-5-612}{\emph{American
  Mineralogist} {\bfseries 86} (2001) 714--730}.

\bibitem{FARGES2004176}
F.~Farges, Y.~Lefrère, S.~Rossano, A.~Berthereau, G.~Calas and G.~E. Brown,
  \emph{{The effect of redox state on the local structural environment of iron
  in silicate glasses: a combined XAFS spectroscopy, molecular dynamics, and
  bond valence study}},
  \href{http://dx.doi.org/https://doi.org/10.1016/j.jnoncrysol.2004.07.050}{\emph{Journal
  of Non-Crystalline Solids} {\bfseries 344} (2004) 176--188}.

\bibitem{WILKE200471}
M.~Wilke, G.~M. Partzsch, R.~Bernhardt and D.~Lattard, \emph{{Determination of
  the iron oxidation state in basaltic glasses using XANES at the K-edge}},
  \href{http://dx.doi.org/https://doi.org/10.1016/j.chemgeo.2004.08.034}{\emph{Chemical
  Geology} {\bfseries 213} (2004) 71--87}.

\bibitem{Tack:ve5108}
P.~Tack, B.~Bazi, B.~Vekemans, T.~Okbinoglu, F.~Van~Maldeghem, S.~Goderis
  et~al., \emph{{Investigation of (micro-)meteoritic materials at the new hard
  X-ray imaging PUMA beamline for heritage sciences}},
  \href{http://dx.doi.org/10.1107/S160057751901230X}{\emph{Journal of
  Synchrotron Radiation} {\bfseries 26} (Nov, 2019) 2033--2039}.

\bibitem{Medjoubi:vv5054}
K.~Medjoubi, N.~Leclercq, F.~Langlois, A.~Buteau, S.~L{\'{e}}, S.~Poirier
  et~al., \emph{{Development of fast, simultaneous and multi-technique scanning
  hard X-ray microscopy at Synchrotron SOLEIL}},
  \href{http://dx.doi.org/10.1107/S0909049512052119}{\emph{Journal of
  Synchrotron Radiation} {\bfseries 20} (Mar, 2013) 293--299}.

\bibitem{SOLE200763}
V.~Sol\'e, E.~Paillon, M.~Cotte, P.~Walter and J.~Susini, \emph{{A
  multiplatform code for the analysis of energy-dispersive X-ray fluorescence
  spectra}},
  \href{http://dx.doi.org/10.1016/j.sab.2006.12.002}{\emph{Spectrochimica Acta
  Part B: Atomic Spectroscopy} {\bfseries 62} (2007) 63--68}.

\end{thebibliography}\endgroup
\end{document}